\documentclass{aastex6}
\usepackage{newtxtext,newtxmath}
\usepackage[T1]{fontenc}
\usepackage{ae,aecompl}

\usepackage{graphicx}	% Including figure files
\usepackage{amsmath}	% Advanced maths commands
\usepackage{amssymb}	% Extra maths symbols
\usepackage{afterpage}

\bibpunct{(}{)}{;}{a}{}{,}
\def\aap{A\&A}                % Astronomy and Astrophysics
\def\ao{Applied Optics}       % Applied Optics
\def\apjl{ApJ}                % Astrophysical Journal, Letters
\def\apjs{ApJS}               % Astrophysical Journal, Supplement
\newcommand{\HII}{H\,{\sc ii}}

\newcommand{\HB}{H{$\beta$}}

\newcommand{\HEI}{He\,{\sc i}}
\newcommand{\HEII}{He\,{\sc ii}}
\newcommand{\HEIII}{He\,{\sc iii}}
\newcommand{\Halpha}{H$\alpha$}

\def\p0{\phantom{0}}

\def\arcmin{\hbox{$^\prime$}}
\def\arcsec{\hbox{$^{\prime\prime}$}}

\def\p0{\phantom{0}}

\begin{document}
%%%%%%%%%%%%%%%%%%% TITLE PAGE %%%%%%%%%%%%%%%%%%%

% Title of the paper, and the short title which is used in the headers.
% Keep the title short and informative.
\title[LMC X--1: A New Spectral Analysis of the O-star in the binary]{LMC X--1: A New Spectral Analysis of the O-star in the binary and surrounding nebula}

% The list of authors, and the short list which is used in the headers.
% If you need two or more lines of authors, add an extra line using \newauthor
\author{E. A. Hyde$^{1}$}
\affil{Western Sydney University, Locked Bag 1797, Penrith, NSW 2751, Australia$^{1}$}
%\affil{Australian Astronomical Observatory, PO Box 296 Epping, NSW 1710, Australia$^{7}$}
\author{D. M. Russell$^{2}$}
\affil{New York University Abu Dhabi, PO Box 129188, Abu Dhabi, United Arab Emirates$^{2}$}
\author{A. Ritter$^{3}$}
\affil{Department of Astrophysical Sciences, Princeton University, 4 Ivy Lane, New Jersey 08544, USA$^{3}$}
\author{M. D. Filipovi\'c$^{1}$}
\affil{Western Sydney University, Locked Bag 1797, Penrith, NSW 2751, Australia$^{1}$}
\author{L. Kaper$^{5}$}
\affil{Astronomical Institute Anton Pannekoek, University of Amsterdam, Science Park 904, P.O. Box 94249, 1090 GE, Amsterdam,
The Netherlands $^{5}$}
 \author{K. Grieve$^{1}$}
\affil{Western Sydney University, Locked Bag 1797, Penrith, NSW 2751, Australia$^{1}$}
\and
 \author{A. N. O'Brien$^{1,6}$}
 \affil{Western Sydney University, Locked Bag 1797, Penrith, NSW 2751, Australia$^{1}$}
\affil{CSIRO Astronomy and Space Science, Marsfield, NSW 2122, Australia$^{6}$}
% List of institutions

%$^{5}$Department of Astronomy and Astrophysics, UCO/Lick Observatory, University of California, Santa Cruz, CA 95064, USA\\
\altaffiltext{1}{e.hyde@westernsydney.edu.au}

\begin{abstract}

We provide new observations of the LMC~X--1 O star and its extended nebula structure using spectroscopic data from VLT/UVES as well as H$\alpha$ imaging from the Wide Field Imager on the Max Planck Gesellschaft / European Southern Observatory 2.2m telescope and ATCA imaging of the 2.1 GHz radio continuum.  This nebula is one of the few known to be energized by an X-ray binary.  We use a new spectrum extraction technique that is superior to other methods to obtain both radial velocities and fluxes. This provides an updated spatial velocity of $\simeq 21.0~\pm~4.8$ km s$^{-1}$ for the O star.  The slit encompasses both the photo-ionized and shock-ionized regions of the nebula. The imaging shows a clear arc-like structure reminiscent of a wind bow shock in between the ionization cone and shock-ionized nebula. The observed structure can be fit well by the parabolic shape of a wind bow shock.  If an interpretation of a wind bow shock system is valid, we investigate the N159-O1 star cluster as a potential parent of the system, suggesting a progenitor mass of $\sim 60$~M$_{\odot}$ for the black hole. We further note that the radio emission could be non-thermal emission from the wind bow shock, or synchrotron emission associated with the jet inflated nebula. For both wind and jet-powered origins, this would represent one of the first radio detections of such a structure.

\end{abstract}

\keywords{
stars: black holes  -- X-rays: binaries -- ISM: jets and outflows -- stars: winds, outflows -- shock waves -- stars: evolution -- (stars:) supergiants}

%%%%%%%%%%%%%%%%%%%%%%%%%%%%%%%%%%%%%%%%%%%%%%%%%%

%%%%%%%%%%%%%%%%% BODY OF PAPER %%%%%%%%%%%%%%%%%%

\section{Introduction}
  The brightest X-ray source in the Large Magellanic Cloud (LMC) is designated LMC X--1, a High-Mass X-ray binary (HMXB). Although LMC X--1 was easily discovered in the X-ray \citep{1969ApJ...155L.143M} it took some time to confirm its optical counterpart \citep{1983IAUC.3791....3H}. This counterpart is now thought to be an O star companion referred to as star 32 \citep{1995PASP..107..145C}. 
   
LMC~X--1 is located in one of the many star-forming regions of the LMC known as N\,159. In N\,159 the system of LMC~X--1 is in a highly ionized area designated N\,159F. The accreting matter from the wind of the massive O star onto the black hole in LMC~X--1 creates its high X-ray luminosity. The strong ultraviolet (UV) emission coupled with the X-ray emission results in this system ionizing He as well as H in the nebula N\,159F surrounding LMC~X--1 \citep{pakullone}. Furthermore, as an HMXB, LMC~X--1 represents a relatively short ($10^{4}$~yr) stage in the evolution of a massive close binary, probing an under-studied parameter space of stellar evolution.
   
The LMC X--1 binary system consists of a massive O star and the strongest X-ray source in the LMC: an accreting $10.9\pm1.4~M_{\odot}$ black hole \citep{2009ApJ...697..573O}. The LMC environment is a metal-poor, gas-rich irregular dwarf galaxy, containing many giant and supergiant stars \citep{2005AJ....129..776N}. The 30~Dor region, some 20\arcmin~north of the N\,159F structure around LMC X--1, is the most active star-forming region in the Local Group. Due to the high star formation rate there are a large number of SNRs in the LMC ($\sim$80).  

The wind of the massive O star in LMC X--1 is not only a source for accretion but also interacts with the interstellar medium (ISM). If the system has a high space velocity it may form a shock where it interacts with the ISM; this is a ``runaway'' system. In the case of HMXB Vela X--1 this resulted in the formation of a shell or a wind bow shock around the system, since it is moving supersonically with respect to the interstellar medium \citep{1997ApJ...475L..37K}. 

The structure of the highly ionized nebula N\,159F around LMC~X--1 \citep{pakullone} is complex \citep{2007ApJ...667L.163C,2008ApJ...687L..29C}. The large, $\sim15$~pc diameter nebula is more than 10 times larger than the Str\"{o}mgren radius for star 32, it is shell-like and visible in many optical emission lines. The relative strengths of the lines favor a shock wave traveling at $\sim90$~km~s$^{-1}$ \citep{2007ApJ...667L.163C}, which these authors interpret as a bow shock powered by the as-yet undetected jet \citep[e.g.][]{2006csxs.book..381F} in LMC~X--1, much like the ring bow shock of Cygnus X--1 \citep{2005Natur.436..819G}. Historically, these jets are observed in radio emission, and the material in the jet can be relativistic \citep[e.g.][]{2006MNRAS.367.1432M,2010MNRAS.404L..21C}. Various radio-continuum surveys \citep{1994PASAu..11...68H,1995A&AS..111..311F,2007MNRAS.382..543H} of this region didn't show any significant emission, partly because of their poor resolution and sensitivity.

It is believed that these relativistic radio jets can energize the ISM, and this has been confirmed by observations of decelerating jets \citep[e.g.][]{2002Sci...298..196C} and by shock waves powered by X-ray binary jets \citep{1998AJ....116.1842D,2005Natur.436..819G,2006MNRAS.372..417T,2009MNRAS.397L...6W,2010Natur.466..209P}. In the case of Cygnus X--1, its jet is thought to have pushed a shock wave out from the binary into the surrounding region, forming a ring-like structure (about 5 pc in diameter) around the system \citep{2005Natur.436..819G}. Such nebulae can be used as calorimeters; the energetics of the nebula imply an average jet power in Cygnus X--1 of $P_{\rm Jet} < 2x10^{38}$erg s$^{-1}$ \citep{2007MNRAS.376.1341R,2015MNRAS.446.3579S}.

The highly photo-ionized nebula bright in He\,{\sc ii} is much smaller and closer to LMC~X--1, and was found to be directional, approximating a 3.8~pc cone that aligns with the direction of the jet implied by the larger bow shock nebula \citep{2008ApJ...687L..29C}. This ionization cone is likely energized by the accretion process onto the black hole in LMC~X--1 \citep{2008ApJ...687L..29C}. No other X-ray binaries are known to possess X-ray ionized nebulae, but some X-ray/UV ionized regions have been found around ultra-luminous X-ray sources \citep[e.g.][]{2004MNRAS.351L..83K,2009ApJ...697..950K,2011ApJ...731L..32M}. Uncertainties in the orbital path and true spatial direction of the LMC X-1 system preclude us ruling out either a pure wind bow shock or a pure jet-powered scenario. In this work we investigate the consistency of the wind bow shock scenario.

Characterizing this system has been an ongoing challenge in astronomy and we add perspective to the system on the following points:
\begin{itemize}
\item{We perform new radio continuum observations at 2.1 GHz around the LMC X--1 region, tracing the shocked gas.}
\item{We use high resolution spectroscopy to confirm the spectral type of the O-star and show some evidence that the spectral type is consistent with Of but we do not rule out a spectral type of Of?p. We additionally provide a table of line measurements for the system.}
\item{We present the evidence for a wind bow shock, fit the parabola and show that the velocity is consistent with the space velocity of the system.}
\item{This wind bow shock, bright in He\,{\sc ii}, can be represented as a separate structure to the large-scale collisionally ionized nebula, the latter of which is most likely energized by the jets.}
\item{We constrain the positional origin of LMC~X--1 using its now refined space velocity, and assess the scenario of its birth within a known Supernova Remnant (SNR).}
\end{itemize}

\section{Observations}
We combine optical and radio observations to investigate the LMC X--1 area. This approach allows us to study the emission from the O star in LMC X--1 and from the shocked gas around the system.

\subsection{VLT/UVES spectrum star 32}
We use a Very Large Telescope (VLT) Ultraviolet and Visual Echelon Spectrograph (UVES) \citep{2000SPIE.4008..534D} commissioning spectrum of the O star companion (Star 32) as well as H$\alpha$ images of the N\,159 region to investigate the environment and learn more about the O star of LMC~X--1. We also use for the first time the new extraction routine of \cite{2014PASP..126..170R} in order to obtain accurate values of both radial velocities and the emission line fluxes from both the O star and the nebula \citep{2004MNRAS.351L..83K}. The lines from star 32 are used to constrain its spectral type, and we use line fluxes and imaging of the nebula to investigate its energetics. Finally we use the results to investigate the potential origins of the binary LMC~X--1 and constrain the progenitor mass of the black hole.

The echelle spectrum was taken during the commissioning of the VLT/UVES instrument on January 12, 2000 (51555 MJD). This instrument was constructed for maximum mechanical stability (positioned in the Nasmyth platform of UT2 Kuyen, Unit Telescope 2 of the VLT) and has a resolving power of $R\simeq 45,000$ in the blue and $R\simeq 43,000$ in the red arm for a 1 arcsec slit. The spectrum is covered by three detectors \citep[one in the blue arm and two in the red arm;][]{2000SPIE.4008..534D}. The standard settings of the VLT/UVES instrument contain two dichroic options. The data presented herein were taken with Dichroic $\# 1 (390+564)$ using central wavelengths at 3900 {\AA}  and 5640 {\AA}. This provides a range of $3300-4500$~\AA\ for the blue arm and $4625-6640$~\AA\ for the red arm. There is a gap in the coverage in the red arm between the two CCDs (due to the physical separation of 0.96 mm of the red arm CCDs), which results in a gap in our wavelength coverage from 5600 to about 5650 {\AA}.

\subsubsection{VLT/UVES Data Reduction}
The first iteration of data reduction was done with the ESO UVES pipeline and GASGANO\footnote{From the ESO website \emph{https://www.eso.org/sci/software/gasgano.html}}. However, due to the severe contamination of the spectrum by the surrounding nebula, the analysis required a thorough extraction procedure. To accomplish this we used the STELLA data-reduction pipeline from \cite{ritter1}. 

\begin{figure*}
\vbox to220mm{\vfil
\centering              
 \includegraphics[width=0.95\textwidth]{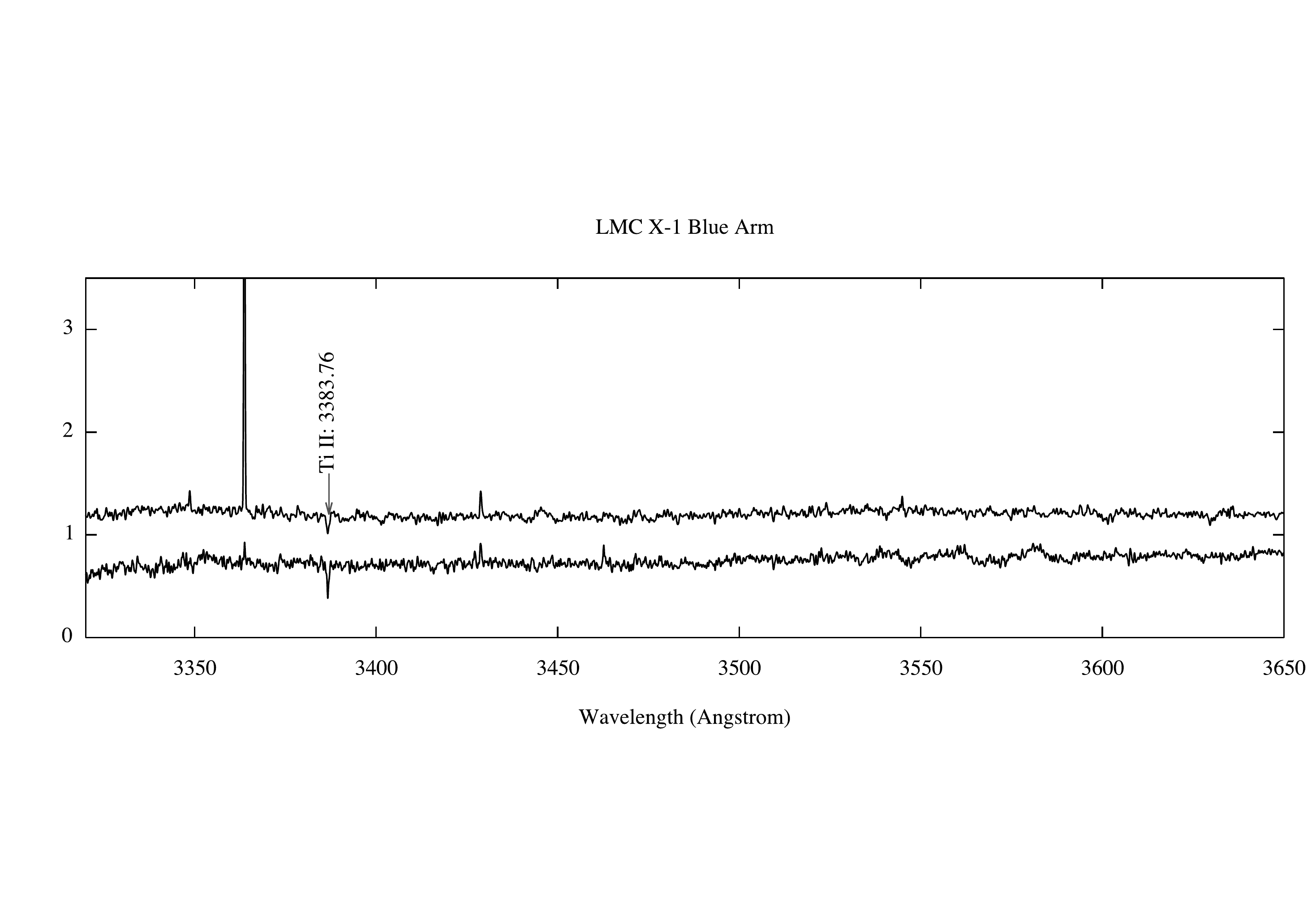} 
  \includegraphics[width=0.95\textwidth]{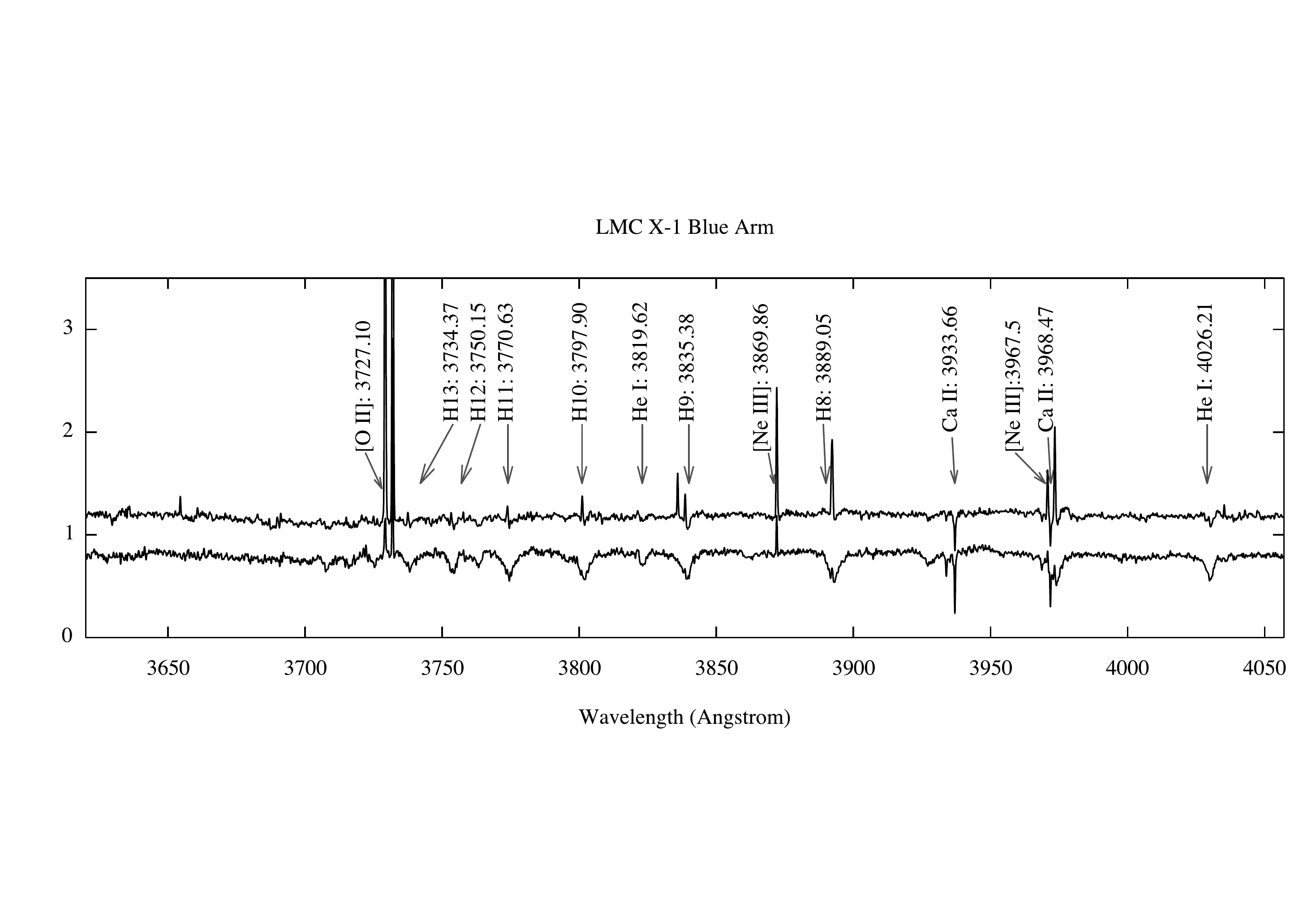} 
   \includegraphics[width=0.95\textwidth]{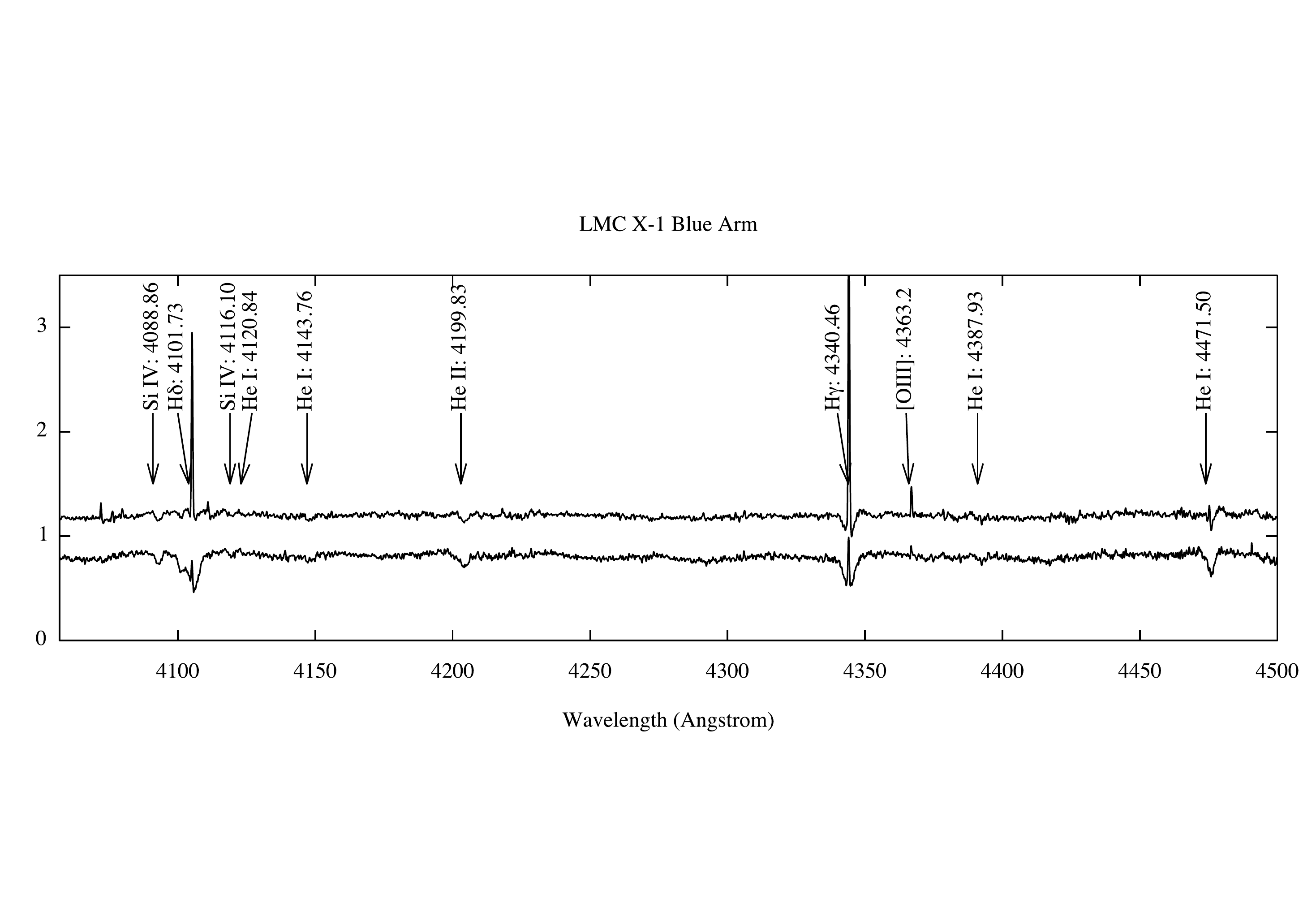} 
    
 \caption{The reduced spectrum of LMC~X--1, first section (blue wavelengths). The original spectrum is shown above and the sky-subtracted spectrum is shown below.}\label{fig:normspec11}
\vfil}

\end{figure*}

\begin{figure*}
\vbox to220mm{\vfil
 \includegraphics[width=0.95\textwidth]{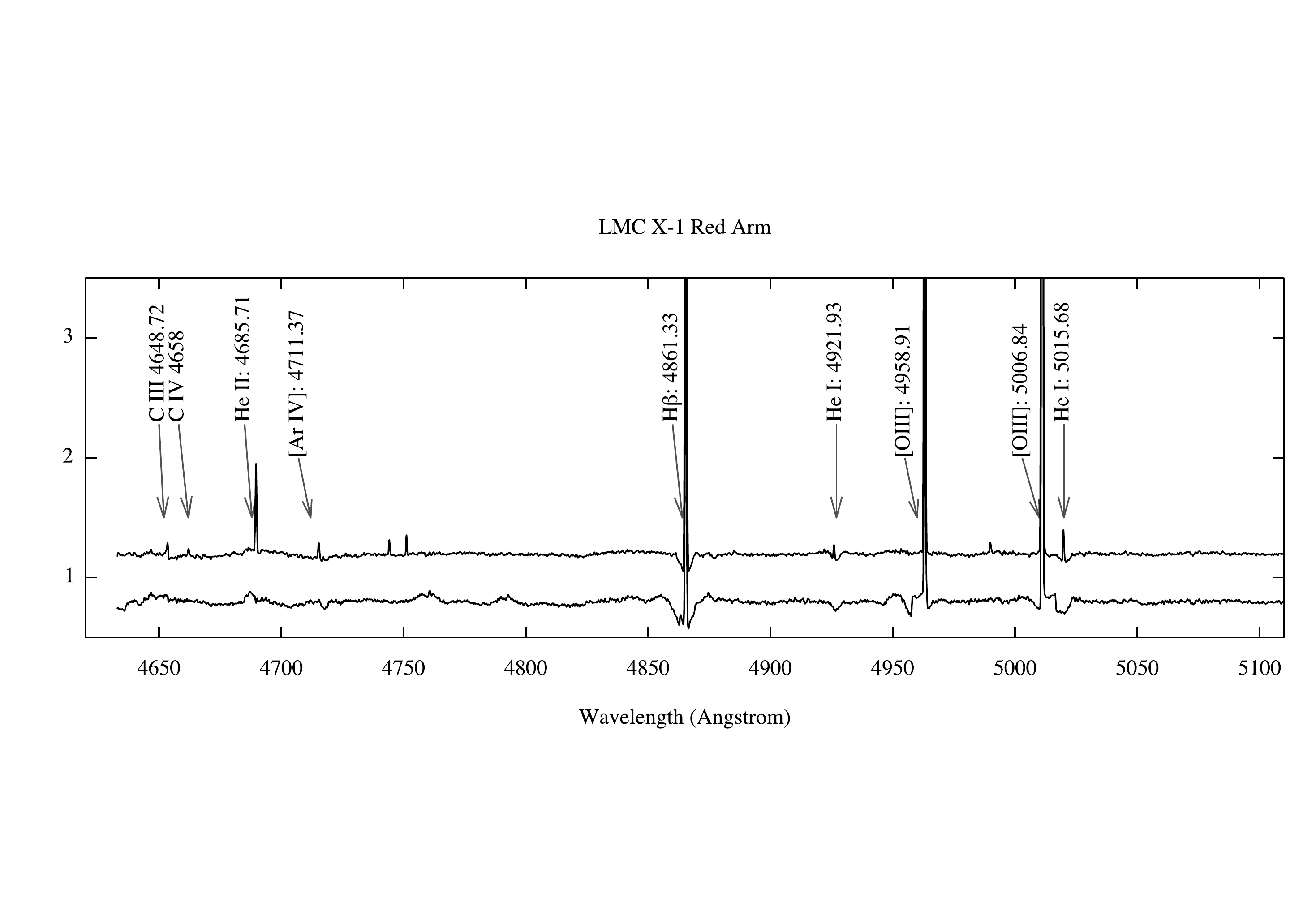}
     \includegraphics[width=0.95\textwidth]{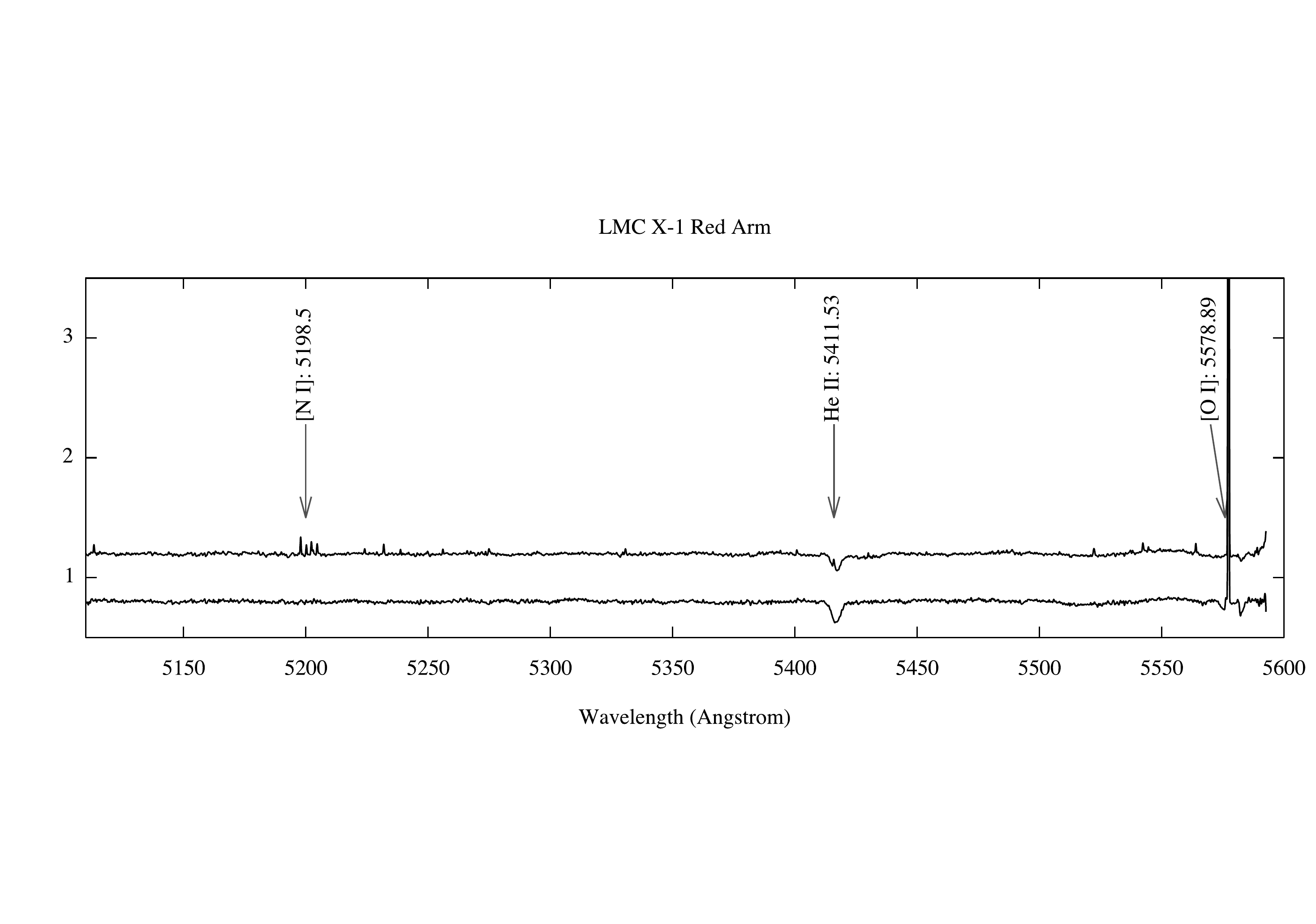} 
      \includegraphics[width=0.95\textwidth]{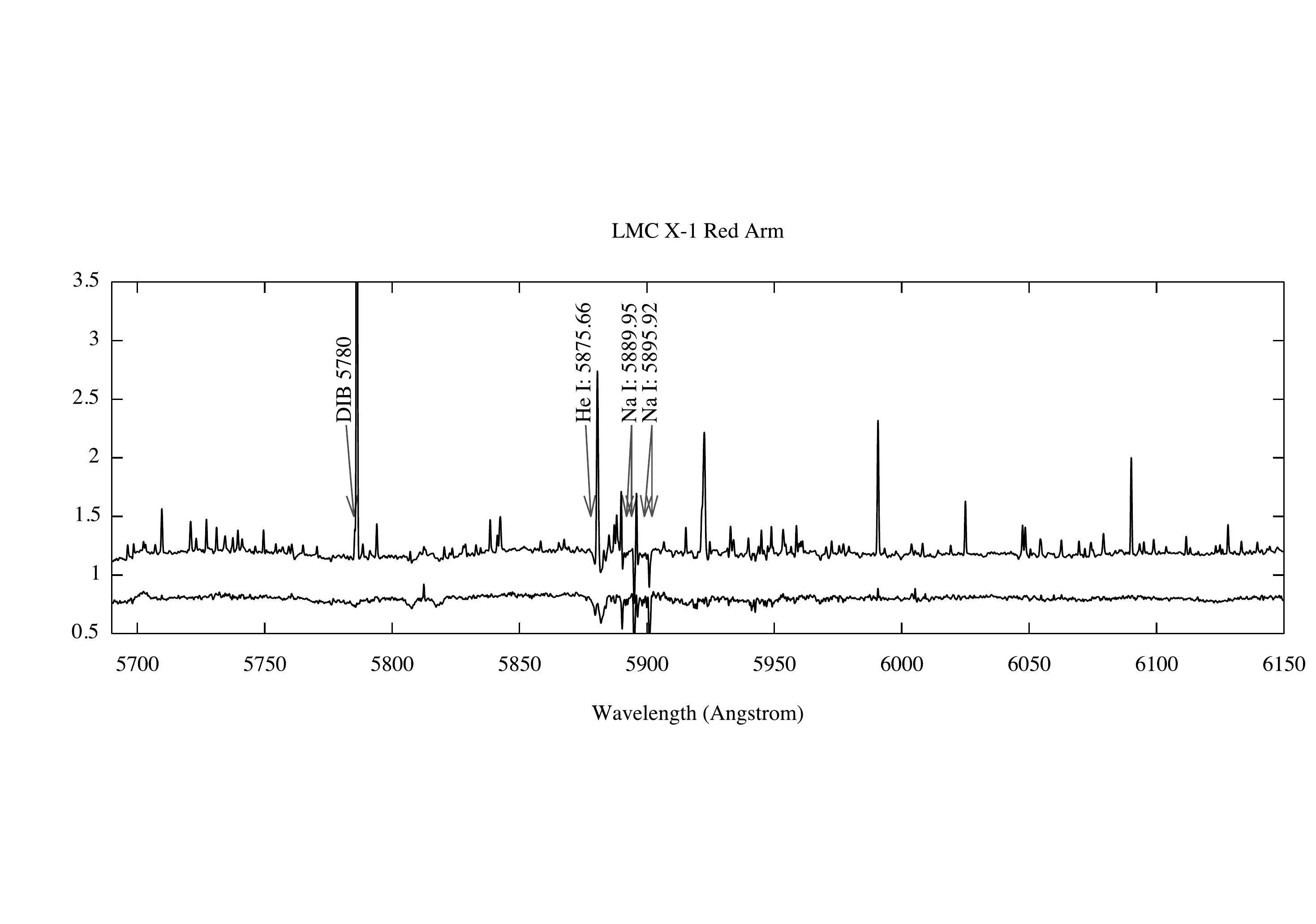} 
       \includegraphics[width=0.95\textwidth]{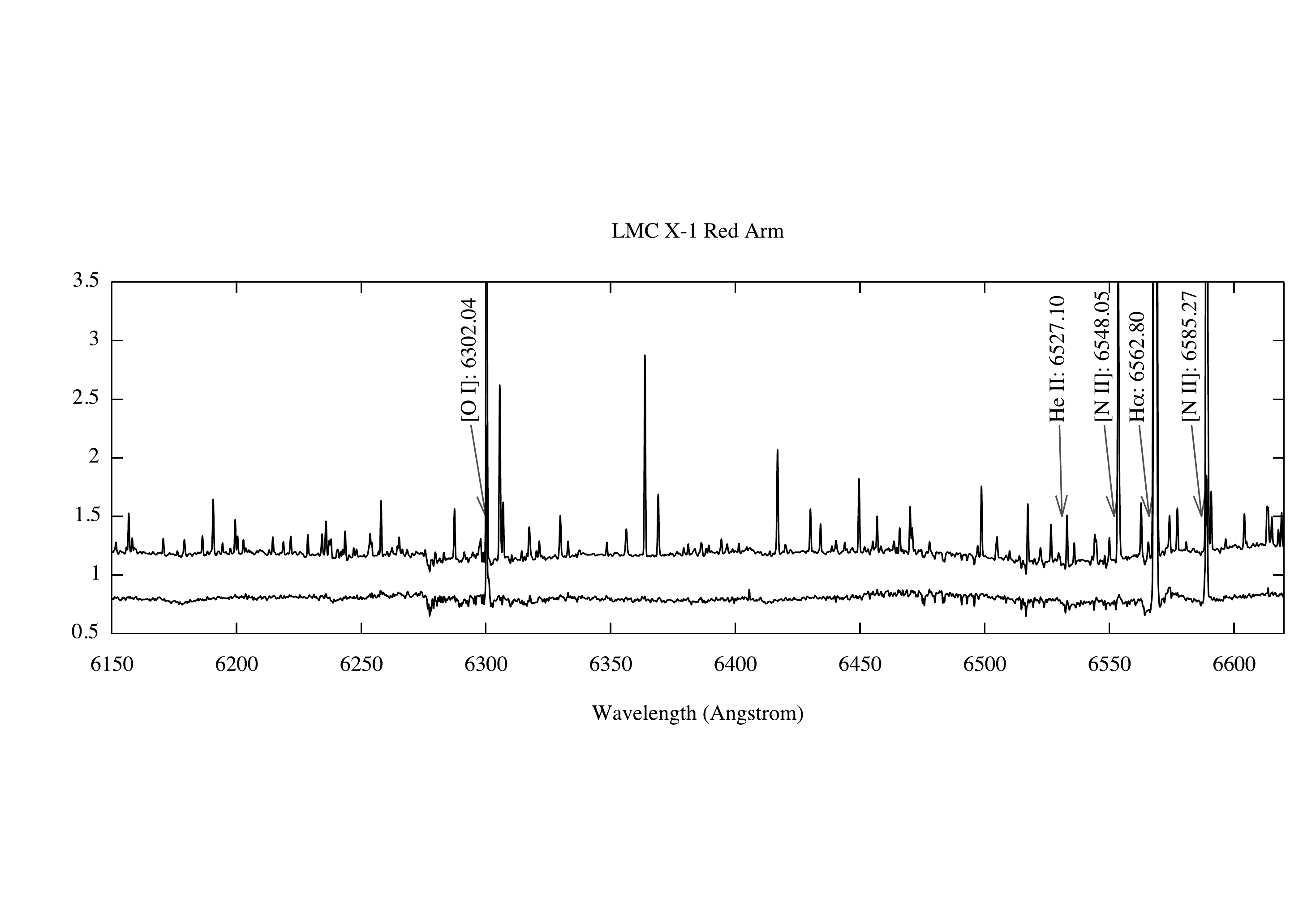} 
\caption{The reduced spectrum of LMC~X--1, second section (red wavelengths). Here again the sky-subtracted spectrum is shown below. We note there is a remnant from the nebular line subtraction near [OIII] 5006.}\label{fig:normspec22}
\vfil}
\end{figure*}

The science frames were overscan and bias subtracted, flat-fielded, scattered-light subtracted, and extracted using a fast and stable re-implementation of the optimal extraction algorithm by \cite{2002AA...385.1095P} in combination with a new optimal sky-subtraction algorithm. This new sky subtraction algorithm is the first that calculates the sky from the whole aperture. It calculates the most probable spatial profile and scales this profile plus a constant to the measured spectrum row by row in an iterative process. For a detailed description please refer to \cite{2014PASP..126..170R}. After sky subtraction and optimal extraction the science spectra were wavelength calibrated, re-binned with oversampling, and merged without jumps (another advantage of the STELLA pipeline). The final spectra are shown in Figures \ref{fig:normspec11} and \ref{fig:normspec22}.

%\clearpage

\subsection{ATCA Radio-continuum observations}
\label{sec:radio}
Our new Australia Telescope Compact Array (ATCA) observations (project CX-310) used the Compact Array Broadband Backend (CABB) with the 6A array configuration at 2.1\,GHz providing improved flux density estimates and resolution compared to previous radio observations. These new images were acquired on January $2^{\mathrm{nd}}$, 2015 (57024 MJD) with $\sim1$ hour integration over the 12-hour observing session. The radio galaxy PKS 1934-638 was used as a primary flux calibration source for all observations, with the radio source PKS~0530-72 used for phase calibration. A standard calibration process was carried out using the \textsc{miriad} data reduction software package \citep{2011ascl.soft06007S}. In order to improve the fidelity and sensitivity of the final image, a single iteration of self-calibration was performed on the strongest sources in the field. Given the 2\,GHz of bandwidth provided by CABB, images were formed using \textsc{miriad} multi-frequency synthesis \citep[\textsc{mfclean};][]{1994A&AS..108..585S}. The same procedure was used for both {\it U} and {\it Q} Stokes parameter maps. 

 There was no reliable detection in the Q or U intensity parameters associated with this object, implying a lack of polarization at lower radio frequencies (2.1 GHz). We report 3$\sigma$ noise level in our Q and U images to be 0.75 mJy/beam. This translates into upper limit for fractional polarization detection at level of $<$5\%. This absence of polarization may be due to bandwidth depolarization which is particularly severe at high fractional bandwidths such as in these observations.

The Stokes I image was created using a Briggs weighting scheme with a robust parameter of zero. This produced a more sensitive image allowing the detection of the extended shock front at the cost of producing a larger synthesized beam. The initial image contained phase calibration errors which resulted in significant artifacts around strong sources within the field. These errors were corrected with a single iteration of phase-only self-calibration using a masked model containing the strong sources produced by the task `MFCLEAN' and a solution interval of 0.1 minutes. This was sufficient to remove the image artifacts and further self-calibration iterations produced noise-like gain corrections and thus did not further improve the fidelity of the image. 

\begin{figure}
 \includegraphics[width=.455\textwidth]{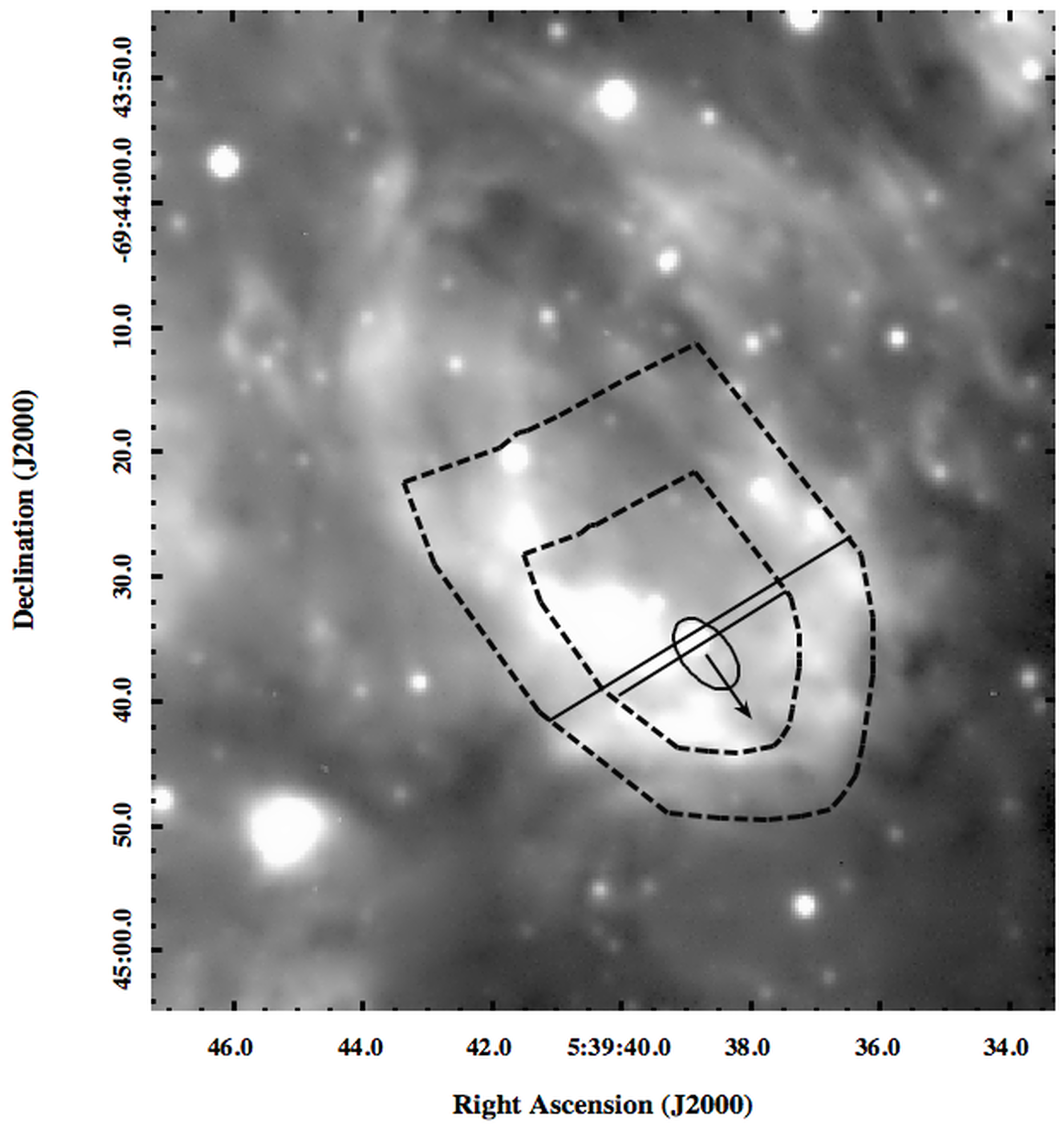}
       \includegraphics[width=0.52\textwidth]{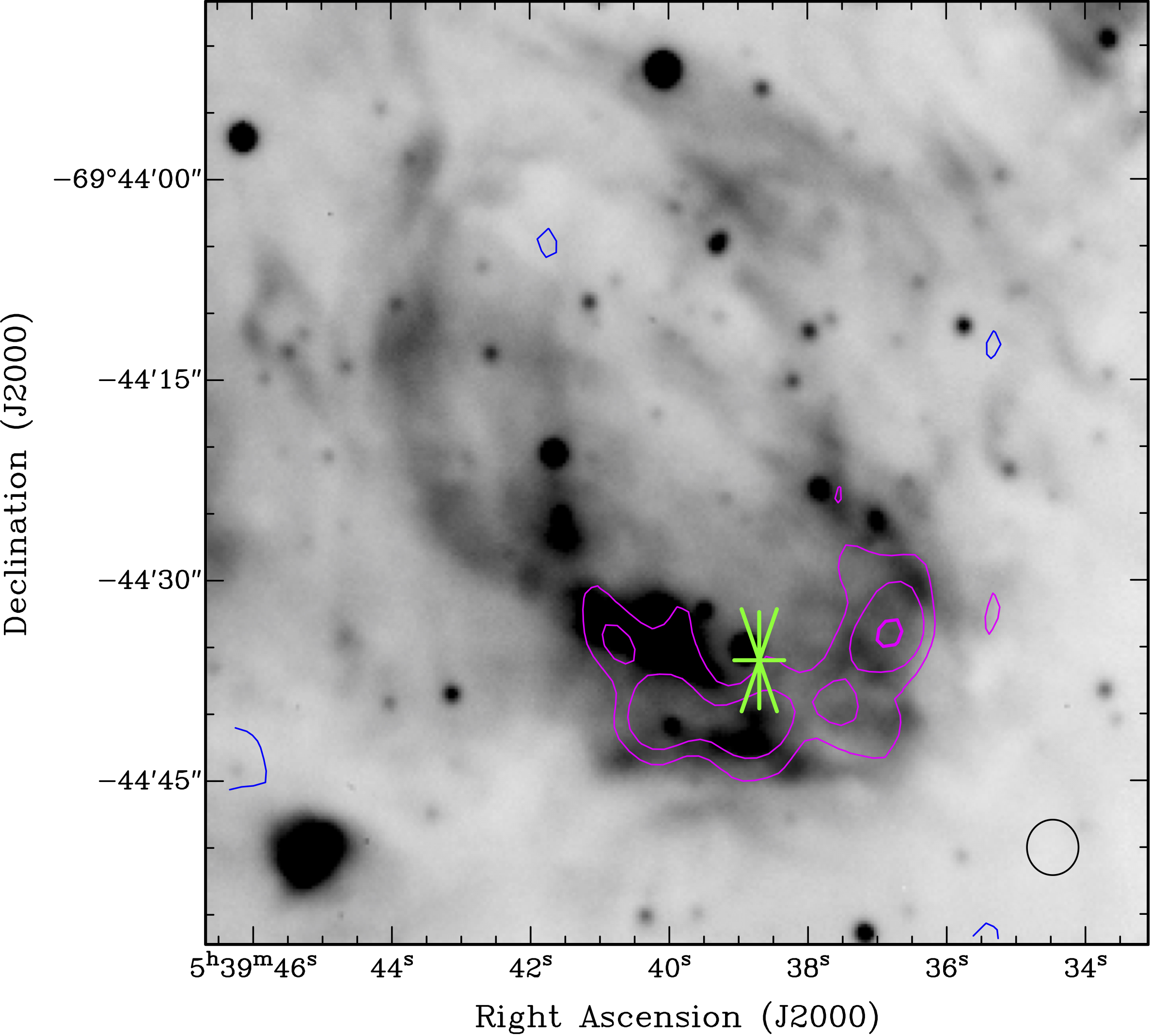} 
\caption{\emph{Left:} The H$\alpha$ image of LMC~X--1 region taken with the 2.2-m ESO/MPI telescope and the Wide Field Imager (WFI) showing a parabolic fit (see Section 4), with inner and outer diameters, for the nebula around LMC X--1. Top diameter lines are shown parallel but measurements are from the center coordinates of the O Star. The ionization cone is shown as an ellipse as estimated by \protect\cite{2008ApJ...687L..29C}. The standoff distance direction is shown by the arrow. The position of LMC~X--1 roughly corresponds to the ellipse in this image.\emph{Right:} The same H$\alpha$ image adjusted and overlaid with ATCA 2.1\,GHz surface brightness image contours (-0.75~mJy/beam in blue and 0.75 (3$\sigma$), 1.5, and 3.0~mJy/beam in magenta). The beam size is given by the black circle. The green star represents the position of LMC~X--1 (the ellipse given in the left image).}
\label{fig:radio}
\end{figure}

The final image produced (after primary beam corrected) had a FWHM of $4.15\arcsec \times 3.86\arcsec$ and a PA of $-0.8^{\circ}$, with a 1$\sigma$ rms (1$\sigma$) noise level of 0.25~mJy/beam. Figure~\ref{fig:radio} shows ATCA 2.1\,GHz surface brightness image contours (-0.75~mJy/beam in blue and 0.75 (3$\sigma$), 1.5, and 3.0~mJy/beam in magenta) of LMC~X--1.

%\afterpage{\clearpage}
%\clearpage

\subsection{WFI optical imaging observations}
\label{sec:wfi}
We have also obtained wide-field imaging observations of the region of LMC~X--1, which were made with the Wide Field Imager (WFI) on the Max Planck Gesellschaft (MPG) / European Southern Observatory (ESO) 2.2m telescope situated at La Silla, Chile in February 2006 under ESO program ID: 076.D-0017(A). The data were taken on February $26^{\mathrm{th}}$, 2006 (53792 MJD). The WFI camera \citep{1999Msngr..95...15B} contains eight 2000 $\times$ 4000 pixel CCDs; its field of view is 34 by 33 arcmin, with a resolution of 0.238 arcsec per pixel. Two 500-sec exposures were made in the H$\alpha$ filter. The conditions were good and the airmass was 1.5. The exposures were dithered to account for the gaps between the chips. Bias exposures and flats were taken on the same nights as the science data. The data reduction package \small THELI \normalsize \citep{2005AN....326..432E} was used to de-bias, flat-field and combine the data. Positional calibration was achieved in \small THELI \normalsize by matching several hundred stars in each exposure with those in the online MAST Guide Star Catalog. The images were then stacked, resulting in one mosaic image. This image is shown as the background in Figure \ref{fig:radio}. 

The morphology of the N\,159F nebula is shown in the 2.1 GHz ATCA image contours as well as the WFI H$\alpha$ image in Fig. \ref{fig:radio}. The morphology of the nebula can be compared to the arc-like structure shown in the \HEII/H$\beta$ imaging of~\cite{2008ApJ...687L..29C}. This shape is roughly parabolic and what we expect in the case of a wind bow shock as discussed in Section \ref{sec:windbowshock}. The WFI optical observations and the contours from our ATCA radio data shown in Fig. \ref{fig:radio} allow us to estimate a graphical fit for a wind bow shock solution in Section \ref{sec:windbowshock}.

\section{Analysis of the Spectral Lines}
\begin{table*}
\centering
{\tiny
\begin{tabular}{| c | c | c | c | c | c | c | c | c | c | c |}
\hline
line &rest $\lambda$ ({\AA}) &obs $\lambda$ ({\AA}) & obs flux($*100/$H$\alpha$) &obs flux& $A_{\lambda}/ A_{\rm V}$& de-red flux($*100/$H$\beta$)&Hvel (km s$^{-1}$) & FWHM& Hvel error (km s$^{-1}$) \\
\hline							
Star and Sky\\
\hline
\hline
 Nebular Lines  \\
\hline
$[$OII$]$            &3726.03           & 3729.18        & 4.995              &2.757       &1.54       &69.37      &253.62   &0.235      &9.46\\
$[$OII$]$		&3728.73		&3731.94		&6.874		&3.794	&1.54	&95.47	&258.26	&0.237	&9.55\\
H13			&3734.37		&3737.52		&0.139		&0.077	&1.53	&1.93	&253.05	&0.451	&18.11\\
H12			&3750.15		&3753.31		&0.143		&0.079	&1.53	&1.97	&252.78	&0.476	&19.03\\
H11			&3770.63		&3773.81		&0.184		&0.1018	&1.53	&2.51	&253.01	&0.518	&20.61\\
H10			&3797.9		&3801.1		&0.238		&0.131	&1.52	&3.19	&252.77	&0.4079	&16.11\\
\HEI			&3819.62		&3821.31		&0.104		&0.057	&1.51	&1.38	&132.73	&0.782	&30.70\\
H9			&3835.38		&3838.63		&0.246		&0.136	&1.51	&3.23	&254.21	&0.3458	&13.52\\
$[$NeIII$]$	&3868.69		&3872.04		&1.200		&0.6624	&1.50	&15.47	&259.78	&0.336	&13.04\\
H8			&3889.049	&3892.21		&1.012		&0.558	&1.49	&12.89	&243.83	&0.604	&23.29\\
\HEI			&4026.191	&4035.24		&0.109		&0.061 	&1.45	&1.28	&674.26	&0.2184	&8.13\\
H$\delta$		&4101.74		&4105.21		&1.893		&1.045	&1.43	&20.98	&253.79	&0.4179	&15.28\\
\HEI$^{*}$			&4143.76		&4147.96		&-0.241		&-0.133	&1.41	&-0.13	&304.07	&2.618	&94.76\\
H$\gamma$	&4340.47		&4344.15		&3.565		&1.968	&1.35	&33.06	&254.35	&0.438	&15.13\\
$[$OIII$]$		&4363.15		&4366.9		&0.302		&0.1669	&1.34	&2.76	&257.84	&0.419	&14.42\\
\HEI			&4471.5		&4475.29		&0.221		&0.122	&1.29	&1.85	&254.27	&0.3951	&13.25\\
CIV			&4659.6		&4662.05		&0.040		&0.02216	&1.23	&0.29	&157.74	&0.328	&10.56\\
\HEII			&4685.74		&4689.66		&0.921		&0.5083	&1.22	&6.57	&250.97	&0.475	&15.22\\
\HEII			&4685.74		&4687.19		&0.783		&0.4322	&1.22	&5.59	&92.83	&4.058	&129.90\\
$[$Ar IV$]$        &4711.3  		&4715.23 		& 0.069  	&0.038  &1.21  &0.48  &250.25  &0.1334  &4.24\\
H$\beta$		&4861.33		&4865.39		&15.812		&8.727	&1.16	&100	&250.54	&0.4875	&15.04\\
\HEI			&4921.9		&4926		&0.138		&0.076	&1.14	&0.84	&249.90	&0.3562	&10.85\\
$[$OIII$]$		&4958.83		&4963.04		&21.834		&12.05	&1.13	&129.68	&254.70	&0.358	&10.83\\
$[$OIII$]$		&5006.77		&5011.01		&65.718		&36.27	&1.12	&379.01	&254.06	&0.364	&10.91\\
\HEI			&5015.7		&5019.85		&0.264		&0.145	&1.12	&1.51	&248.22	&0.3325	&9.94\\
\HEII			&5411.43		&5415.92		&0.137		&0.075	&1.02	&0.63	&248.91	&0.5216	&14.45\\
$[$OI$]$		&5577.26		&5577.34		&12.658		&6.986	&0.98	&54.69	&4.30	&0.101	&2.72\\
\HEI			&5875.6		&5880.58		&2.073		&1.144	&0.93	&7.98	&254.27	&0.481	&12.27\\
$[$OI$]$		&6300.2		&6305.57		&1.475		&0.814	&0.86	&4.91	&255.71	&0.206	&4.89\\
$[$OI$]$		&6363.67		&6369.1		&0.509		&0.2809	&0.85	&1.66	&255.98	&0.210	&4.94\\
$[$NII$]$		&6547.96		&6553.57		&2.857		&1.577	&0.82	&8.78	&257.03	&0.332	&7.61\\
H$\alpha$	&6562.8		&6568.33		&100		&55.19	&0.82	&305.57	&252.79	&0.685	&15.65\\
$[$NII$]$		&6583.34		&6588.96		&8.400		&4.636	&0.81	&25.48	&256.10	&0.395	&8.99\\
\hline
O Star Line\\
\hline
CIII			&4648.72		&4653.51		&0.094		&0.0523	&1.23	&0.70	&309.12	&0.333	&10.74\\

\hline
Sky Subtracted\\
\hline	
\hline
Interstellar Lines\\
\hline		                                                                                                           
TiII			&3383.76		&3386.68		&-0.246	&-0.136	&--&		-0.135	&258.88	&0.2848	&12.63\\
CaII(MW)		&3933.66		&3933.88		&-0.154	&-0.085	&--&		-0.084	&16.78	&0.3263	&12.44\\
CaII			&3933.66		&3937		&-0.385	&-0.213	&--&		-0.211	&254.72	&0.2397	&9.14\\
CaII(MW)		&3968.47		&3968.72		&-0.129	&-0.071	&--&		-0.071	&18.90	&0.55	&20.79\\
CaII			&3968.47		&3971.84		&-0.247	&-0.136	&--&		-0.135	&254.76	&0.2235	&8.45\\
DIB5780		&5780		&5785.02		&-0.760	&-0.420	&--&		-0.417	&260.55	&6.711	&174.16\\
NaID(MW)	&5889.95		&5890.31		&-0.284	&-0.157	&--&		-0.156	&18.34	&0.3845	&9.79\\
NaID			&5889.95		&5894.98		&-0.641	&-0.354	&--&		-0.352	&256.20	&0.3164	&8.06\\
NaID(MW)	&5895.92		&5896.35		&-0.216	&-0.119	&--&		-0.118	&21.88	&0.3872	&9.85\\
NaID			&5895.92		&5900.94		&-0.513	&-0.283	&--&		-0.281	&255.43	&0.2808	&7.14\\
\hline
Nebular Lines\\
\hline
H13			&3734.37		&3738.24		&-0.826	&-0.456	&--&		-0.453	&310.90	&3.627	&145.69\\
H12			&3750.15		&3753.85		&-0.890	&-0.491	&--&		-0.488	&295.99	&2.662	&106.48\\
H11			&3770.63		&3774.42		&-1.527	&-0.843	&--&		-0.838	&301.54	&4.048	&161.03\\
H10			&3797.9		&3801.79		&-1.749	&-0.965	&--&		-0.959	&307.28	&4.113	&162.45\\
\HEI			&3819.62		&3823.22		&-0.394	&-0.218	&--&		-0.216	&282.75	&1.835	&72.06\\
H9			&3835.38		&3839.01		&-1.581	&-0.872	&--&		-0.867	&283.94	&3.999	&156.39\\
H8			&3889.049	&3892.96		&-1.841	&-1.016	&--&		-1.010	&301.69	&4.195	&161.80\\
SiIV			&4088.862	&4092.94		&-0.415	&-0.229	&--&		-0.228	&299.20	&2.072	&76.01\\
\HEI$^{*}$			&4143.76		&4145.33		&-2.640	&-1.457	&--&		-1.448	&113.66	&23.08	&835.47\\
\HEI			&4713		&4717.7		&-0.420	&-0.232	&--&		-0.230	&299.17	&2.914	&92.74\\
\HEI			&4921.9		&4926.92		&-0.532	&-0.294	&--&		-0.292	&305.98	&3.800	&115.81\\
\HEII			&5411.43		&5416.92		&-1.454	&-0.803	&--&		-0.798	&304.36	&4.285	&118.78\\
\hline
%Other Lines\\
%\hline

\hline
\end{tabular}}
\caption{Spectral lines from LMC X--1 as measured from the non-sky subtracted VLT/UVES spectrum. All rest wavelengths and FWHM values are in Angstroms ({\AA}) and are taken in air as given by the NIST atomic database. The ``Hvel'' parameter is the heliocentric velocity measured. The de-reddened fluxes adopt the extinction value of $A_{\rm V} = 2.28 \pm 0.06$ \protect\citep{2009ApJ...697..573O} and the extinction law of \protect\cite{1988ESASP.281b.215C}. The errors are propagated from the uncertainty in the extinction only, not the error in the UVES line flux measurements. The \HEI$^{*}$ 4143 line is measured twice as the sky subtraction appears to have interfered with the absorption line.}\label{t:measvel}
\end{table*} 

We convert the wavelength positions of the detected lines into radial velocities using the \small IRAF \normalsize task \small SPLOT\normalsize, and record the line radial velocity measurements in Table \ref{t:measvel}. The errors in the \small SPLOT \normalsize measurements are given in the routine from a Poisson model of the pixel sigmas. This approximately corresponds to the 1$\sigma$ errors for that particular line or, in the case of radial velocities, an average error of $\pm 2.5$ km s$^{-1}$ for all measurements.

The radial velocity recovered has five main components: the velocity that is due to the motion of the Earth around the sun; the velocity due to the binary motion of Star 32 in its orbit;  the velocity of the LMC with respect to our Galaxy; the velocity of the solar system around the Galaxy and a component that is the spatial velocity of LMC~X--1 with respect to the LMC. Since it is the last value we are interested in, a few important corrections need to be applied to our measurements. We convert to a heliocentric velocity using the IRAF \emph{rvcorrect} command, which gives us a correction of $-0.2$ km s$^{-1}$ for our data. This is extremely small compared to the correction for the orbital phase.

The motion of the O star around the binary orbit was also subtracted from the velocity by analyzing the orbital phase of the binary at the time of measurement. The orbital phase is the fractional part of $(T-T_{\rm o})/P$~\citep{SphereA}. We know that the orbital period is $P = 3.90917 \pm 0.00005$ days and $T_{\rm o} = 2453391.3436 \pm 0.0080$ (in HJD) is defined such that $T_{\rm o} + nP$, for an integer n, gives the phase at periastron from~\cite{2009ApJ...697..573O}. We use the fraction of the total difference in phases and find $\Phi = 0.39$ (roughly the same for both observations) as the phase during our observations. Using the radial velocity curve from Figure 6 of Orosz et al. this phase corresponds approximately to a radial velocity of $300-305$ km s$^{-1}$. The system velocity is about 260 km s$^{-1}$, or, a correction of 40$-$45 km s$^{-1}$ to each value measured. Using a system velocity of 260 km s$^{-1}$ and graphically matching the phase from~\cite{2009ApJ...697..573O} we find an overall correction of $-40.2$ km s$^{-1}$ to each star radial velocity value measured.\footnote{Note that the average velocity of LMC~X--1 is not exceptional for its location \citep{malaroda}.}

\begin{figure}
%\centering
     \includegraphics[width=0.48\textwidth]{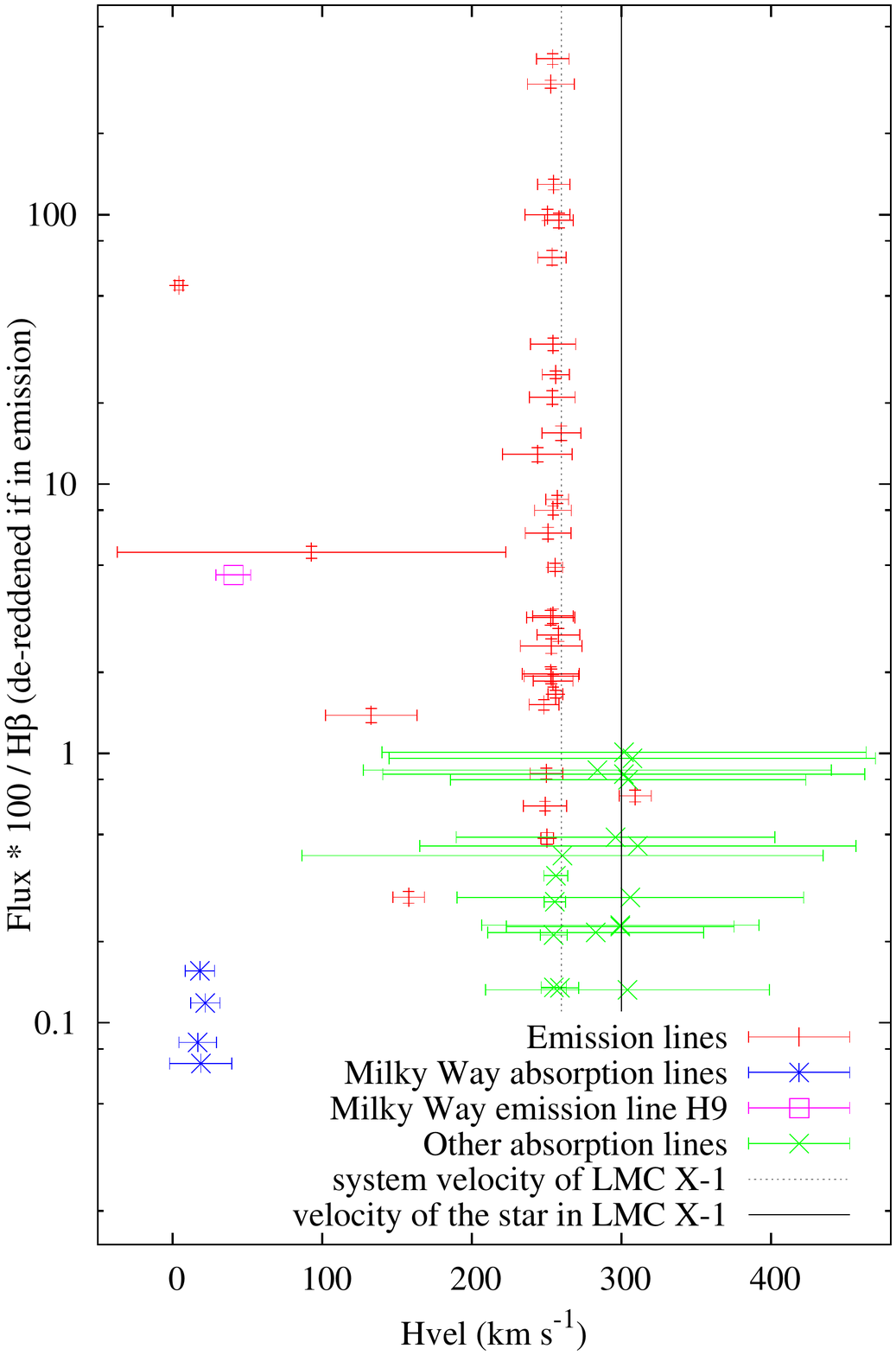} 
      \includegraphics[width=0.48\textwidth]{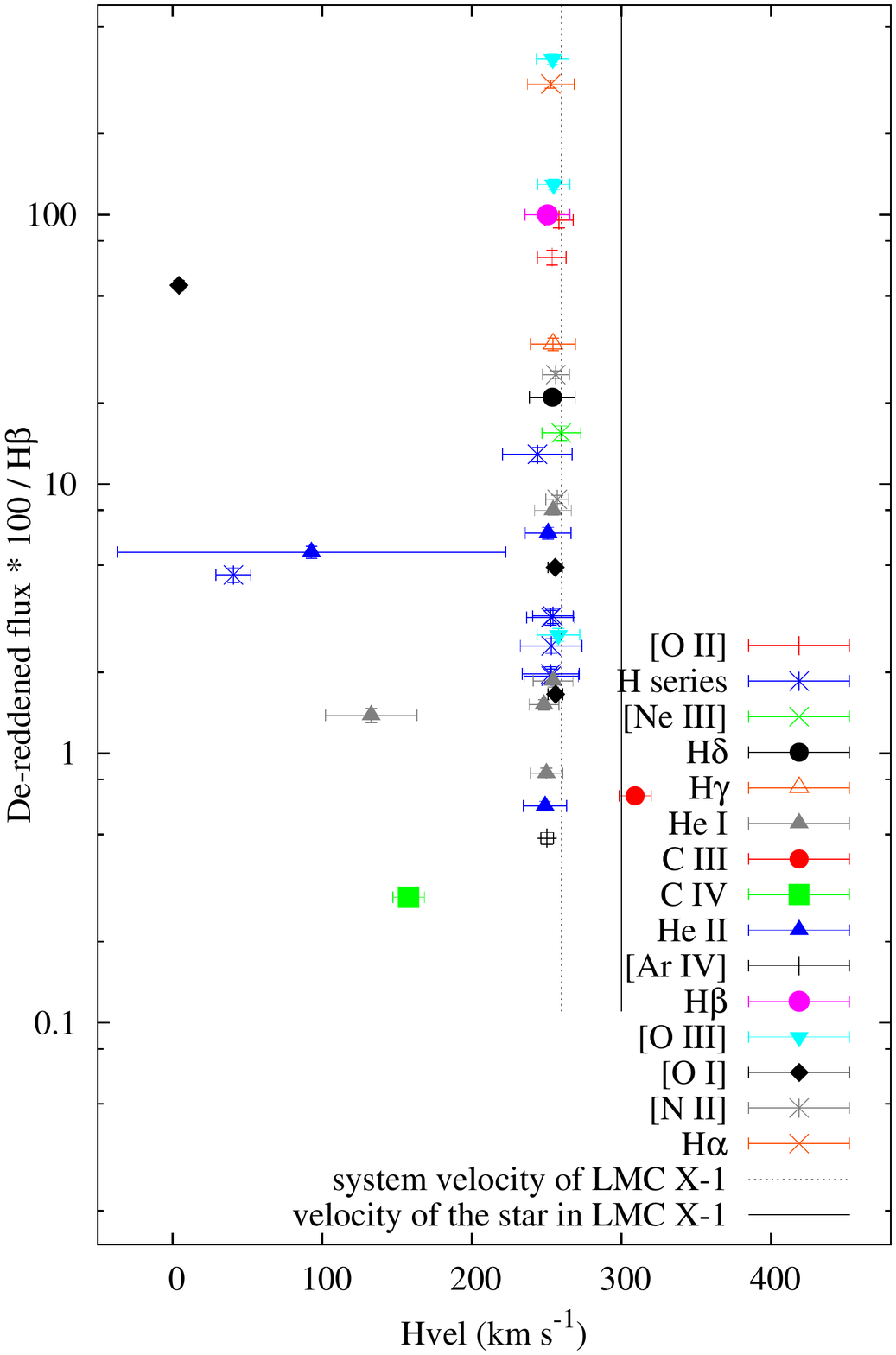} 
\caption{\emph{Left:} Radial velocity versus the de-reddened relative flux of emission lines (red plusses and purple squares) and absorption lines (blue stars and green crosses). \emph{Right:} The same as the left panel but only the emission lines, with the symbols indicating the line identifications. In both panels, the vertical dotted line at 260 km s$^{-1}$ is the system velocity of LMC X--1, and the solid black line at 300 km s$^{-1}$ is the velocity of the star in LMC X--1 at the time of the VLT/UVES observation.}

\label{fig:lines-vel-flux}
\end{figure}

In Fig. \ref{fig:lines-vel-flux} the observed flux of each emission line is plotted against its radial velocity, as shown in Table \ref{t:measvel}. The nebula lines are visible at $\sim 260$ km s$^{-1}$, centered around the velocity of the LMC~X--1 system.

\subsection{Literature versus Measured: Spectral Features of Star 32}
\label{sec:spectraltype}

In Fig. \ref{fig:normspec11} and \ref{fig:normspec22} we see emission lines from the N\,159F nebula as well as many stellar absorption lines and interstellar lines. The emission lines shown in Figure \ref{fig:lines-vel-flux} indicate that the velocity of the emission components are at approximately a radial velocity of $\sim$260~km~s$^{-1}$; very close to the system radial velocity of LMC~X--1.

\begin{figure}
\centering
     \includegraphics[width=0.49\textwidth]{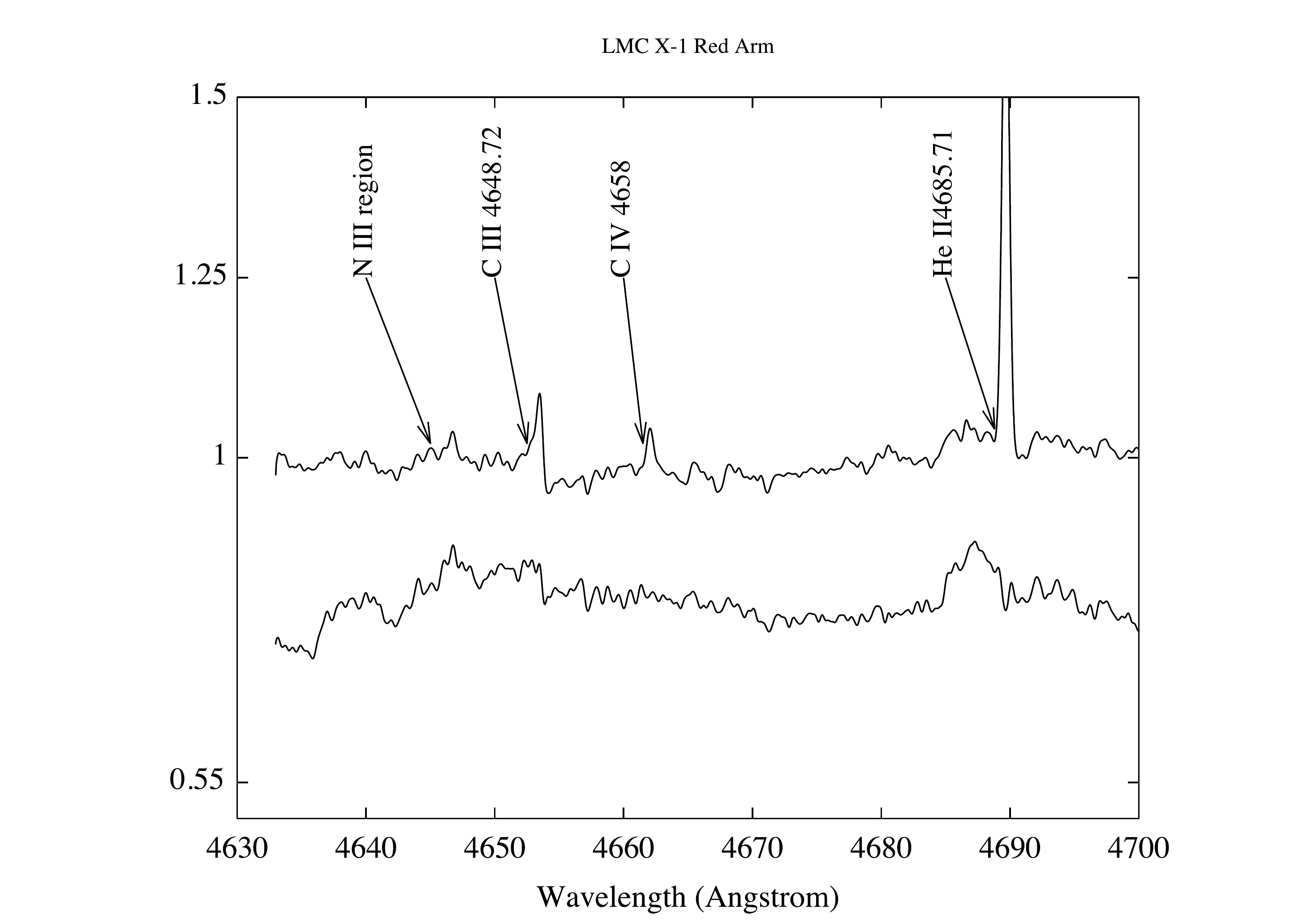} 
       \includegraphics[width=0.48\textwidth]{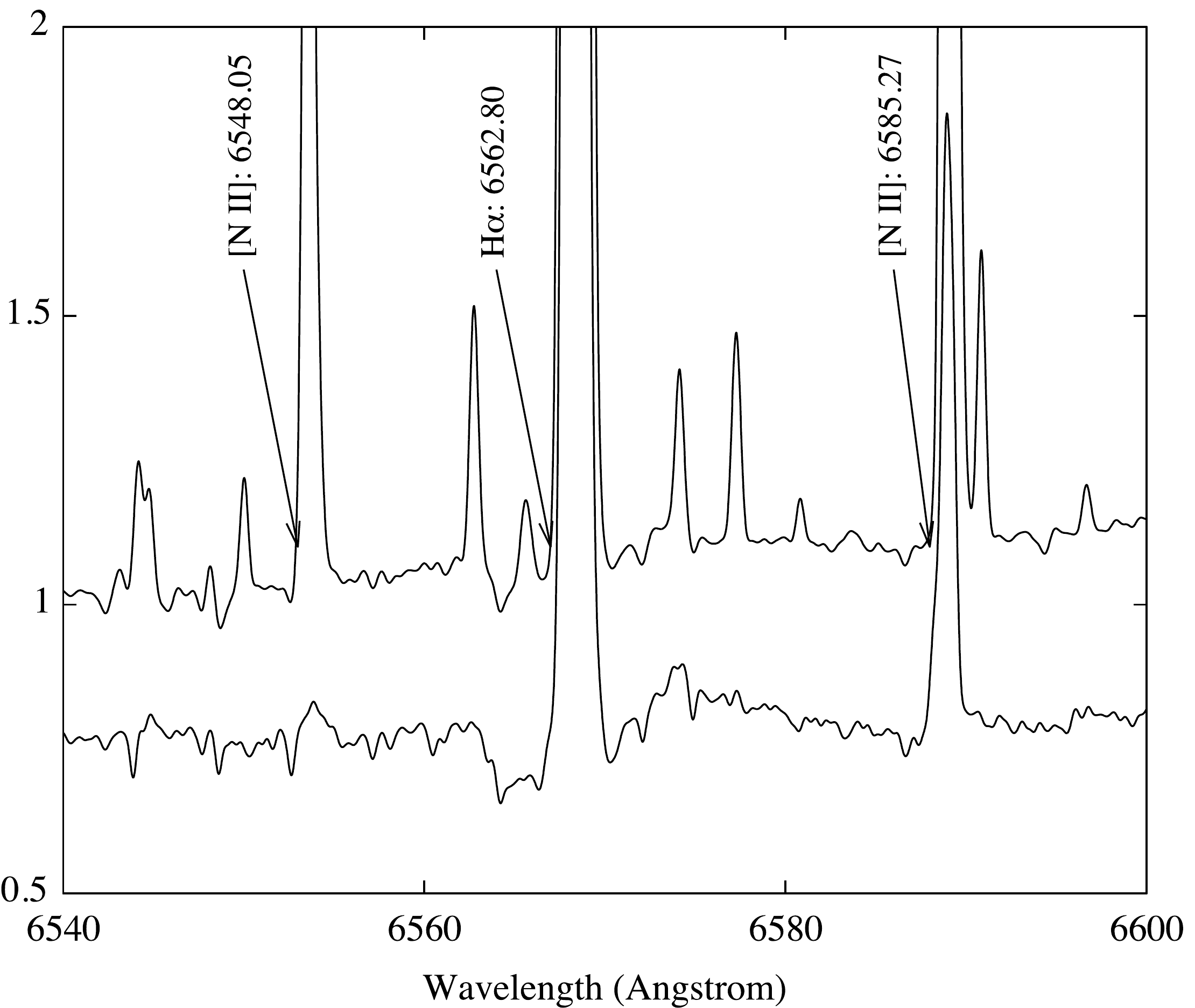} 
\caption{\emph{Left:} The enlarged region of the reduced spectrum of LMC~X--1 around 4650 to 4700 {\AA}. In this plot we can see very high \HEII~emission and the N III emission complex, a characteristic feature for Of classification, is weakly present. C III and C IV emission lines are also present. \emph{Right:} The area around the H$\alpha$ emission, which shows potentially a P-Cygni profile. In both cases the top is the original and the bottom is the sky subtracted spectrum.}
\label{fig:spectype}
\end{figure}

In Fig. \ref{fig:spectype}, enlarged regions of the spectrum around the \HEII~4686 and H$\alpha$ 6563 emission lines are presented. The H$\alpha$ absorption line may have a P-Cygni type profile, although it is difficult to identify due to the large emission line at the same location. The presence of the P-Cygni profile is important because it is an indicator of a strong outflow of material from the O star. This outflow most probably takes the form of a strong stellar wind that drives the X-ray emission from the system via accretion onto the black hole.

Several attempts have been made to classify the O star in LMC~X--1. The \HEI, \HEII, Si IV, and Mg II lines were used by \cite{1983ApJ...275L..43H} to show that the star was consistent with an O7 type, but \cite{2002AA...385..517N} updated this when they used \HEI, \HEII, and Mg to show that the O star was more consistent with an O8 III type star. This would mean that it is in the O8 subclass, and it has a surface temperature of about $T \sim 33,000$ K and a luminosity classification III, which corresponds to a normal giant star. For an O8 type star we expect certain line ratios for some lines, as given by \cite{1990PASP..102..379W}. 

In particular, for an O8 star the H$\gamma$ 4340 line should be about 2~{\AA} in equivalent width, \HEI~4471 should be about 0.8~{\AA} and \HEII~4541 should be about 0.6~{\AA} \citep{JJ}. In the case of LMC X--1 this is complicated by the fact that we also have emission from the nebula at H$\gamma$, and \HEII~4541 is outside the covered wavelength range. Our measurements are shown in Table \ref{t:measvel}. We find a FWHM for H$\gamma$ 4340 of 0.438~{\AA} and a FWHM for \HEI~4471 of 0.395~{\AA}. As this does not agree with \cite{JJ} we note that there may be some sky emission feature over-subtraction in our case.

We note that for a spectral type of O8 III we get a corresponding effective temperature of about $35,000$ K from \cite{1989ApJS...69..527H} to $33,000$ K as shown by \cite{2006ApJ...638..409H}.

Looking at the spectrum shown in Fig. \ref{fig:spectype} we note strong \HEII~4686 emission. We also have some emission at approximately 4650 and 4660 {\AA}. We suspect the relatively bright OII lines from this region to be contributing, and the situation is complicated by the nebula in the region. The large number of nebula emission lines means that it is also possible that this is in fact from C III 4650 and C IV 4658 nebula emission. 

There has been some N III emission in the 4640--4643 {\AA} range, noted by \cite{pakullone}, but it is weak, and we do not detect it here. Due to the depleted N III region the pure f classification parameter is unlikely. This classification parameter is relevant for stars with an extended atmosphere that affects the luminosity \citep{JJ}. The enhanced C III emission is mentioned in \cite{2010ApJ...711L.143W} as a marker for Galactic Of?p stars. The Of?p star classification was introduced by~\cite{1972AJ.....77..312W} and indicates doubt that the star in question is a normal supergiant. These stars can lack supergiant wind profiles in the ultraviolet and are thought to be braked, oblique rotators~\citep{2010ApJ...711L.143W}. Asides from the N III region, the spectrum for LMC~X--1 is roughly consistent with an Of classification but we cannot rule out Of?p. The P-Cygni type profile we would expect in an Of?p classification is less than what we see in Figure \ref{fig:spectype}, but what we see could be partly due to background subtraction effects, so we mention it as a possibility only.

\subsection{The Extinction}

We attempt to estimate the extinction towards LMC~X--1 from our spectra using two methods. The first method we employ uses the fluxes measured in Table \ref{t:measvel}. Using our values for H$\alpha$ and H$\beta$ we can derive an independent measure of the extinction towards our source. From~\cite{2010MNRAS.405.1349R} the extinction constant for the LMC is given by $c(H\beta)=[log(H\alpha/H\beta) - log(2.86)]/0.34 $ and using our measured ratio we find $c(H\beta)\simeq 0.6287$, where the extinction constant $c(H\beta)$ is the logarithmic extinction of H$\beta$. To find the total extinction we follow~\cite{2010MNRAS.405.1349R} and use the relation provided by~\cite{1983MNRAS.203..301H}, $E(B-V)=0.689\times c(H\beta)$, together with the definition of extinction, $A_{\rm V}=R_{\rm V}\times E(B-V)$, where in the case of the LMC we can adopt the value of $R_{\rm V}=3.1$ as it has been shown that there is little difference in the value of $R$ compared to galactic values at these wavelengths \citep[from ][]{1983MNRAS.203..301H}. This means that we find $A_{\rm V}\simeq 1.343$.

Another method allows us to use the Diffuse Interstellar Band (DIB) measurements.~The DIB strength increases with reddening and the DIB at 5780 {\AA} can be used to estimate the extinction \citep[e.g.][]{2005AA...436..661C}. The presence of a 5780 {\AA} DIB requires strong UV radiation, and it is thought that the 5780 {\AA} DIB reaches its maximum strength only in regions with a stronger UV field. We measure the 5780 {\AA} DIB line with the \small IRAF \normalsize routine \small SPLOT \normalsize and find a central wavelength of 5785.71 {\AA} and an equivalent width of 0.1225 {\AA}.

By using our measured value of equivalent width (EW) we find that this corresponds approximately to a value of $\log {\rm EW}(5780 DIB) = \log(122.5 {\rm m\AA}) = 2.088$ from the linear relation of~\cite{2006AA...447..991C}. This gives us $\log {\rm E(B-V)} = -0.6$, or extinction of $A_{\rm V} \sim 0.78$. We suspect some contamination from the bright emission line next to the 5780 DIB location as \cite{2009ApJ...697..573O} constrained the extinction to $A_{\rm V} = 2.28 \pm 0.06$, and rule out extinctions of $A_{\rm V} \leq 1.74$.  Taking into account the range of values and the equivalent width measurements of emission lines we suspect that there might be significant errors in the DIB method. Likewise, the fluxes from our spectra may reflect a region too close by to the O star. The estimate of $A_{\rm V}$ from~\cite{2009ApJ...697..573O} uses not only spectroscopy but also photometry and a dynamical model to arrive at a value of $A_{\rm V} = 2.28 \pm 0.06$ for the extinction. Considering the low error and multiple types of estimates associated with their measurement, we adopt the value of~\cite{2009ApJ...697..573O} ($A_{\rm V} = 2.28 \pm 0.06$) for this paper.

\subsection{Spectral lines of N149F}

LMC~X--1 is situated in the highly ionized N\,159F nebula. This region is an \HEIII\ region. \cite{pakullone} used line strengths to identify and characterize ionized regions around LMC~X--1. The emission lines that are detected in our UVES spectrum are listed in Table \ref{t:measvel}. The slit used by Pakull \& Angebault covered a large area of N\,159F, and from this large area they were able to measure line strengths at two different strips in the nebula. 

The work by~\cite{pakullone} identified a \HEIII\ region 3 pc in radius, a \HII\ Str\"{o}mgren sphere at $r=5$ pc with a density $N_{\rm e}=40$ ${\rm cm}^{-3}$ as well as an extended envelope of neutral H visible in 21 cm radiation around the source. They found an X-ray luminosity of $L_{\rm x} = 10^{39}$ erg s$^{-1}$ for LMC~X--1, where the X-ray luminosity is high enough to completely ionize the wind of the O star. This means that the N\,159F nebula consists of a \HEIII\ region (the X-ray ionized nebula), which is embedded in a conventional \HII\ region excited by the O star (at the time of the publication of Pakull \& Angebault, it was thought to be an O7 type star). This in turn is all enclosed within a neutral H region. The X-ray ionized nebula has now been shown to be directional \citep{2008ApJ...687L..29C}. Further outwards from LMC~X--1, the temperature likely continues to decrease and molecular hydrogen, ${\rm H}_{2}$, should be present although it is very difficult to detect (we do know that to the south of LMC~X--1 is the giant molecular cloud N\,159 East). 

We investigate the regions of neutral hydrogen and ionized hydrogen further by looking at the sound speed that these regions will have. Neutral hydrogen begins to become ionized at temperatures of 3000 K, and increasing to 10,000 K we can create an \HII\ region. The ionization cone is shown as the white ellipse in Figure \ref{fig:radio} and corresponds to the region from~\cite{2008ApJ...687L..29C}.

Therefore the \HII\ region at about 5 pc would have a temperature around 10,000 K, using the isothermal sound speed $c_{\rm s} \approx 10\times \sqrt{T/10^{4}K}=10$ km s$^{-1}$. We also know the electron temperature for the inner part of the nebula found by \cite{1985AA...146..242B}, $T_{\rm e} = 12,000~K$, and we get $c_{\rm s} = 10.9$ km s$^{-1}$ as an upper limit to the sound speed inside the nebula. However, although we know that for regions of neutral hydrogen the temperature must be cooler than 10,000 K we do not have any measurements of the neutral hydrogen temperature, so we can only say that the sound speed would have to be less than 10 km s$^{-1}$.

In our spectral line measurements we had a smaller slit than the one used by Pakull \& Angebault. The slit that Pakull \& Angebault used covers most of the N149F nebula, so it was more than 30" across. Our slit was only 12" and it covers only the area of the nebula next to Star 32. In effect, we are measuring over a smaller area. This being the case, we may be measuring only a small part of the shocked gas (in the case of a wind bow shock this is the gas marked out by the parabolic shape in Figure \ref{fig:radio}) and we may have a large component of emission from the ionized regions. Investigating one emission line that we know is due to nebular emission we will identify the velocity structure there.

\begin{figure}  
\centering
  \includegraphics[width=.5\textwidth]{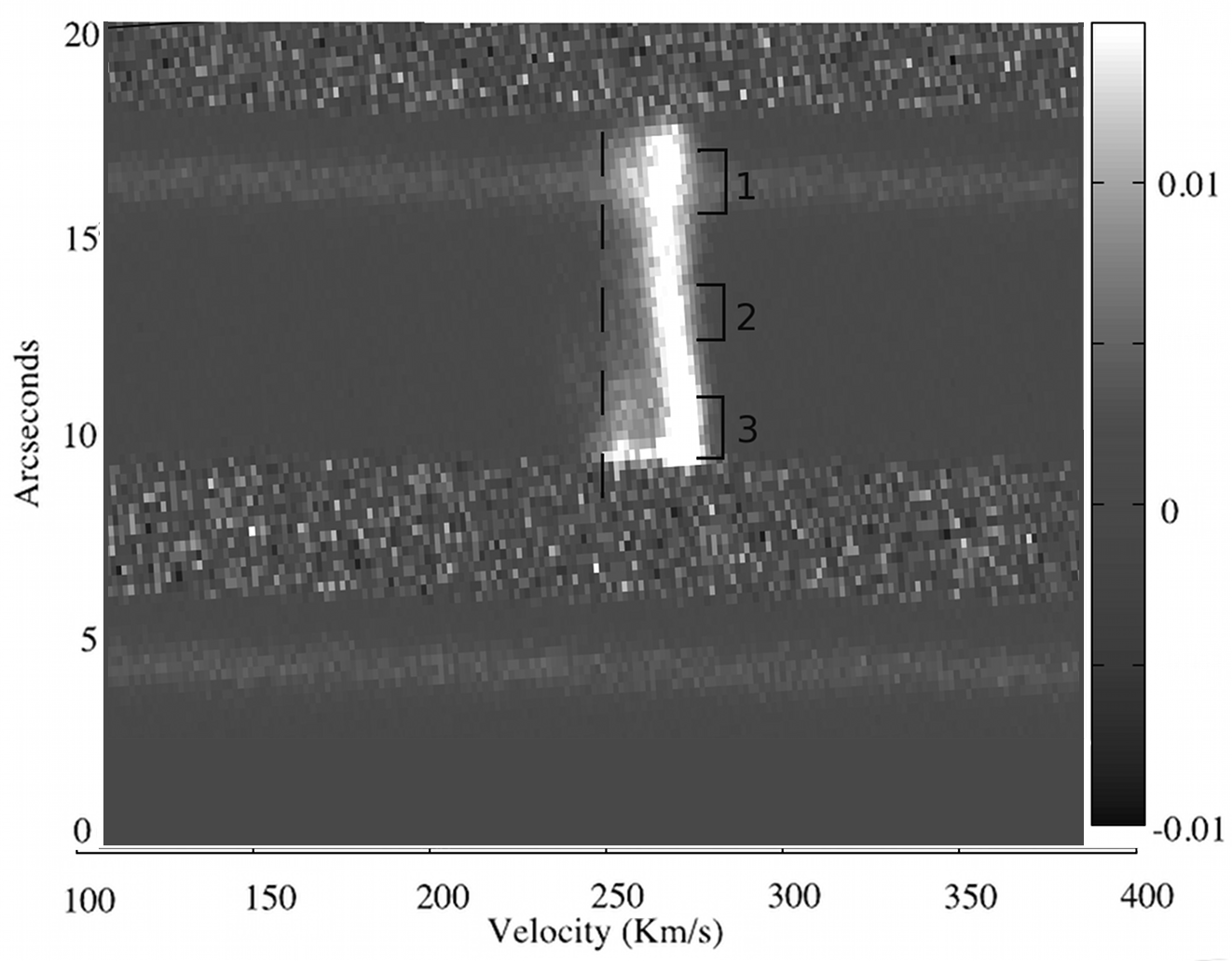}     
 %includegraphics[width=0.5\textwidth]{figures/OII2dvelimage.pdf} 
\caption{The nebular $[$OII$]$ emission line of LMC~X--1 shown in the raw VLT/UVES spectrum. The perpendicular dashed line is at a 90 degree angle to the dispersion and provides a reference for the change in velocity to the emission line. The three marked positions show where we measured the change in velocity of the line. The broad base of the emission line ($\Delta v \approx 50$ km s$^{-1}$) indicates a velocity structure in the nebula.}
 \label{fig:2dOspec}
\end{figure}

Shown in Fig. \ref{fig:2dOspec} is the segment of the echelle spectrum of LMC~X--1 showing $[$OII$]$ 3728. We have converted the image in Fig. \ref{fig:2dOspec} to wavelength scale using the approximate wavelength range of each order shown based on the instrument specifications given by ESO. Using a black body of temperature 36000 K with a V magnitude 14.8 (approximately the values of the O star), order number 125 covers the wavelength range of 3701.6 to 3759 {\AA}, which includes $[$OII$]$ 3728. The conversion is approximate because we have assumed that the orders are not tilted with respect to the columns of the CCD when in fact they are, but over this small range of wavelength the difference is negligible.

There is a gradient that we see in the velocity of the nebula from Figure \ref{fig:2dOspec}, in particular the bottom end of the line is spread in the velocity direction. That spread is almost 50 km s$^{-1}$, which would be greater than the sound speed $c_{\rm s}$ in the nebula. For an electron temperature $T_{\rm e} = 12,000$ K we have an isothermal sound speed of $c_{\rm s} = 10.9$ km s$^{-1}$. As the gradient is greater than the sound speed on the small scale seen through our slit, a shocked flow of material is indicated.

The structure of this nebula is not symmetric (see Fig.\ref{fig:radio}). It has been proposed that this nebula is created by jets \citep{2007ApJ...667L.163C}, and we will also look at the possibility of a wind bow shock caused by the supersonic motion of the system through the ISM.

\section{Evidence for a wind bow shock}
\label{sec:windbowshock}

In this section we will investigate the possibility that part of the nebula surrounding LMC~X--1 is a classical wind bow shock. Towards that goal we will first discuss the nebulosity in the vicinity of LMC~X--1, then wind bow shocks in general and how a velocity estimate of the star can be obtained. 

If the LMC~X--1 system is moving through the interstellar medium (ISM) of the LMC, then from the viewpoint of the star, it is being hit with the ISM particles. The stellar wind balances this pressure, and, if the velocity of the star with respect to the ISM is supersonic, a shock is formed. The location of the shock is determined by the momentum balance between the wind and the ISM.

The wind bow shock around the HMXB Vela X--1 \citep{1997ApJ...475L..37K} is created by a very high space velocity. In particular, the {\it Hipparcos} observations have provided measurements showing that HMXBs can have on average $\sim 50$ km s$^{-1}$. This agrees with previous theoretical work by \cite{2000AA...364..563V}, who predicted runaway velocities in binary evolution scenarios. However, we do not observe wind bow shocks around all HMXBs because in order to create a shock they must be moving faster than the sound speed of the ISM. In the LMC the density of the ISM can be much higher than the average in our Galaxy, and it should then be easier to form a wind bow shock with a lower ISM sound speed \citep{2002AA...383..999H,1998ASPC..131..427K}. We estimate a sound speed of 10 km s$^{-1}$ in the \HII~regions of the LMC and potentially much less than 10 km s$^{-1}$ in the neutral H regions. This implies that if our star does have the high space velocity of some other HMXB systems it will create a shock, depending on the density of the wind and ISM. 

Vela X--1 is one of the most studied wind accreting systems. It consists of a massive BO.5Ib star and a neutron star X-ray pulsar companion. The period is 8.9 days and the parameters of the system were well determined by the pulse timing of the X-ray pulsar \citep{1989PASJ...41....1N}; the BO.5Ib star has a mass of approximately 23 $M_{\odot}$ and the neutron star is about 1.77 $M_{\odot}$. This gives a semi-major axis for the system of $a=0.247$ AU assuming a circular orbit. This compares to the separation of the system of LMC~X--1, where $a=0.179$ AU.

One of the first interesting points of comparison between Vela X--1 and the LMC~X--1 system is the X-ray luminosity. Vela X--1 has an X-ray luminosity of $L_{\rm x}=1.4\times 10^{36}$ erg s$^{-1} = 3.59 \times 10^{2}~L_{\odot}$, which is much lower than the $L_{\rm x} =5 \times 10^{4}$ $L_{\odot}$ of LMC~X--1. This is a typical wind accretion luminosity and a clue that there is additional activity \citep[perhaps some small Roche lobe overflow as suggested by][]{2009ApJ...697..573O} occurring in LMC~X--1, which has a 10 $M_{\odot}$ black hole rather than a $\sim 1.4 M_{\odot}$ neutron star companion in Vela X--1.

Another issue that complicates our binary system is the compact accretor. The accretion radius is where the kinetic energy of the gas is approximately equal to the potential energy of the accretor; in order to calculate this accretion radius we first look at this energy balance, $E_{\rm kin}\simeq E_{\rm pot}$
%\begin{eqnarray}
Since we have steady flows, the time derivatives are zero in the equation of motion and that gives,
\begin{eqnarray}
 \frac{1}{2}m_{\rm gas}v_{\rm gas}^{2} \simeq \frac{GM_{\rm x}m_{\rm gas}}{R_{\rm acc}}
\end{eqnarray}
from \cite{1944MNRAS.104..273B}, where $R_{\rm acc}$ is the accretion radius (in a wind accretor this means where the gravitational force of the compact object is stronger than the wind's kinetic energy). 

This then gives,
\begin{eqnarray}
 R_{\rm acc}\simeq \frac{2GM_{\rm x}}{v_{\rm gas}^{2}}
\label{eqn:racc}
\end{eqnarray}
where $v_{\rm gas}^2 = v_{\rm rel}^{2} + c_{\rm s}^{2}$. We adopt the value for the mass of the black hole as $M_{\rm x} = 10.91$ $M_{\odot}$ \citep{2009ApJ...697..573O} and in this case $c_{\rm s} << v_{\rm rel}$ so we can neglect the sound speed $c_{\rm s}$. 

Next we need to solve for the relative velocity $v_{\rm rel}^{2} = v_{\rm rad}^{2} + v_{\rm tan}^{2}$ of the black hole to the gas, where the velocity components are in the radial and tangential directions respectively. In this case $v_{\rm rad} = v_{\rm w} = v(a) = 748$ km/s from above for the case of Bondi-Hoyle type accretion\footnote{We note that this estimate is different from the literature value of $v_{\rm wind}=1.17\times 10^{6}$ m/s from~\cite{2009ApJ...697..573O} as here we are calculating only the component that will be used for accretion.}, however, since the O star is rotating while the wind is leaving its surface there is an additional tangential component to its velocity. This tangential component is given as $(R_{\rm Star32}/a)v_{\rm rot}$ and combined with the orbital velocity $v_{\rm orb}$, this gives $v_{\rm tan} = v_{\rm orb} - (R_{\rm Star32}/a)v_{\rm rot}$. Since $v_{\rm orb}$ is simply retrieved from the period and Kepler's law, this gives $v_{\rm orb}= (2\pi a)/P = 4.73 \times 10^{7}$ cm/s $= 473$ km/s. Furthermore from \cite{2009ApJ...697..573O} we have an estimate of the radius $R_{\rm Star32} = 17.0 \pm 0.8$ $R_{\odot}$, the inclination $i = 36.38 \pm 1.92 ^{\circ}$, and the rotational velocity $v_{\rm rot}\sin i = 129.9 \pm 2.2$ km/s so the component $(R_{\rm Star32}/a)v_{\rm rot} \simeq 102$ km/s and $v_{\rm tan} \simeq 370$ km/s. Now we solve the relative velocity $v_{\rm rel}^{2} = v_{\rm tan}^{2} + v_{\rm w}^{2} = 7.5 \times 10^{5}$ ${\rm km}^{2}/{\rm s}^{2}$ or $v_{\rm rel} = 834.5$ km/s. Substituting these values in to Equation \ref{eqn:racc} we find that the accretion radius is then $R_{\rm acc} = (2GM_{\rm x})/(v_{\rm rel}^{2}) = 4.16 \times 10^{6}$ ${\rm km} \simeq 5.98$ $R_{\odot}$.

The mass loss rate due to the wind of the O star is also known as the wind density $\rho(a)$. As this is the density at a radius equal to roughly the semi-major axis ($r=a$) we can find this wind density with:
\begin{eqnarray}
\rho(r)\sim\frac{\dot{M_{o}}}{4\pi r^{2}v(r)}
\end{eqnarray}
where $\dot{M_{o}}=\dot{M_{w}}=4.52\times10^{-7}~M_{\odot}/{\rm yr}$ from the formula given by \cite{2001AA...369..574V} and $r=a=36.51~{\rm R}_{\odot}$. Using the radius of the O star $R_{Star32}=17.0\pm0.8~R_{\odot}=1.18\pm0.06\times10^{12}$ cm and at infinity $v_{\inf}=1400$ km s$^{-1}$ we get a velocity $v(a)=v_{\inf}(1-\frac{R_{Star32}}{a})=748\pm31$ km s$^{-1}$ which means at the black hole we have $\rho(a) \simeq 4.7 \times 10^{-15}$ ${\rm g~cm}^{-3}$.

Now we can finally estimate the mass accretion rate via the Bondi-Hoyle formula \citep{1944MNRAS.104..273B}:
\begin{eqnarray}
\dot{M}_{\rm acc} = \pi R_{\rm acc}^{2} \rho(a) v_{\rm rel} = \frac{4\pi G^{2}M_{\rm x}^{2}\rho(a)}{v_{\rm rel}}
\label{eqn:bondi1}
\end{eqnarray}
and by putting in the solved values above, we get that $\dot{M}_{\rm acc} = 2.13 \times 10^{17}$ g s$^{-1}$.

Since we have a black hole as the compact object, its radius is approximately (for a non-spinning black hole) the Schwarzschild radius: 
\begin{eqnarray}
 R_{\rm sch} = \frac{2GM_{\rm x}}{c^{2}}.
\end{eqnarray}
For a mass of $M_{\rm x} = 10.91$ $M_{\odot}$ we get that $R_{\rm sch} = 3.22 \times 10^{6} {\rm cm} = 4.6 \times 10^{-5}$ $R_{\odot}$. For a non-rotating black hole the inner edge of the accretion disk should be at $3\times R_{\rm sch}$. This corresponds roughly to the innermost stable circular orbit (ISCO)~\citep{2013LRR....16....1A}. The accretion luminosity produced will then be the luminosity that was created while the material fell in from the outermost radius to the ISCO. The luminosity of the disk can then be written as $L_{\rm acc}=GM_{\rm x} \dot{M_{\rm acc}}/2\times ISCO$~\citep{2003ApJ...597..780E}.

Solving for luminosity we find that:
\begin{eqnarray}
L_{\rm acc}=\frac{GM_{\rm x} \dot{M_{\rm acc}}}{6\times R_{\rm sch}}\times \approx 1.59 \times 10^{37} {\rm erg/s}.
\label{eqn:L1}
\end{eqnarray}

This luminosity from mass accretion, $L_{\rm acc}= 1.59 \times 10^{37}$ erg/s $\approx 4.1\times10^{3}~L_{\odot}$, is less than the X-ray luminosity of $L_{\rm x} =5 \times 10^{4}$ $L_{\odot}$ from \cite{1986Natur.322..511P}. Pakull \& Angebault derive $L_{\rm x}$ from the degree of ionization of the nebula N\,159F (measured values for the luminosity require $f_{\rm x}$ and the distance d). The difference in our measurement no doubt is partially due to the much larger slit size used by \cite{1986Natur.322..511P}. For this system it is also useful to compare these luminosities with the critical Eddington luminosity of
\begin{eqnarray}
L_{\rm Edd} \simeq 3.3 \times 10^{4}\left[\frac{M}{M_{\odot}}\right] L_{\odot} = 3.6 \times 10^{5} L_{\odot}
\end{eqnarray}
using the mass of the black hole from above. Both our estimate and that of \cite{1986Natur.322..511P} are lower than $L_{\rm Edd}$. In addition the X-ray luminosity of $L_{\rm x}$ is higher than the estimated luminosity but below the critical Eddington luminosity. This indicates that the black hole is possibly accreting below the Eddington limit or further than the estimated ISCO to the black hole.

We see that the typical wind accretor Vela X--1 has a larger semi-major axis and a (generally) smaller accretion radius than LMC~X--1. The mass accretion rate from \cite{1980ApJ...238..969D} is $\dot{M}\sim 10^{-11} M_{\odot}~{\rm yr}^{-1} = 6.3 \times 10^{14}{\rm g~s}^{-1}$, which is about two orders of magnitude lower than what we see in LMC~X--1. So although these systems both show a visible bow shock structure, the LMC~X--1 system has a luminosity that appears rather high.

The fact that LMC~X--1 also has its compact object closer to the massive star and that the massive star is thought to be close to filling its Roche lobe \citep{2009ApJ...697..573O}, might provide a possible explanation for the higher luminosity of the system, but the high X-ray luminosity may also be explained by stalling of the wind due to an ionization zone (the Str\"{o}mgren sphere) around the black hole, allowing matter to accumulate and resulting in a larger $R_{\rm acc}$.

In general for all wind bow shock systems it is expected that where the wind of the massive star interacts with the ISM a shock will be formed in the direction of the motion of the star. This shock can be modeled by a parabola as shown in Figure 1 from \cite{1996ApJ...459L..31W}. This same parabola is reproduced in our Figure \ref{fig:radio} as the graphical structure fit around the nebula surrounding LMC X--1. Using the geometry of the shock we can measure how far the star is from the shock front. This distance from the star to the shock is determined by the momentum balance of the wind and the ISM and is known as the standoff distance. Figure \ref{fig:radio} shows observations of LMC~X--1 with points marked for a parabolic fit, with an inner and outer diameter marked with lines. The diameter lines are shown parallel but measurements discussed are from the center coordinates of the O Star. 

Following the discussion given by \cite{1996ApJ...459L..31W} we can solve the shape of the parabola for the standoff distance, $R_{\rm o}$, by using the geometry of the system. Finding $R_{\rm o}$ is relatively easy for this configuration, defining a distance $D=2R$ and adopting $\theta=90^{\circ}$ we find: 
\begin{eqnarray}
 R_{\rm o}=\frac{D}{2\times \sqrt{3}}
 \label{eqn:Wilkin}
\end{eqnarray}
from the equations given by \cite{1996ApJ...459L..31W}. We will use this geometry and what we know about wind bow shocks to investigate whether this is a viable mechanism for creating the shape we see in the nebula N\,159F that surrounds LMC~X--1 in the following section.

We show that the structure close to LMC~X--1 (shown in Figure \ref{fig:radio}) can be reasonably explained by a wind bow shock. We can use the shape of the shock to constrain the standoff distance. This is done by a graphical fit of a parabola onto the image of the system, as shown in Fig. \ref{fig:radio}. This fit is done by marking points along the visible edge of the bow shock and then fitting a parabolic function to those points, we note that the marking of points is done via a line-up by visual inspection. We use the same fit for both the left and right in Fig. \ref{fig:radio} to show the comparison across the two different images of the same nebula area around LMC X--1. We measure $D$ across the parabola at $\theta=90^{\circ}$ where $D$ passes through the coordinates of the O star. This measurement is independent of the orientation to the system, i.e the projection on the sky of the parabolic shape has a constant value of $D$. 

Converting the measurements of $D$ that we derive from the image (adopting 0.266 pc arcsec$^{-1}$ at a distance of 55 kpc), using Equation \ref{eqn:Wilkin}, we constrain $R_{\rm o}$ as shown in Table \ref{t:Ro}. Cooke et al. (2008) have shown that the X-ray ionized gas is mostly on one side of the LMC~X--1 system, (this, the ionization cone, is marked in Fig. \ref{fig:radio}). The outer boundary of this cone appears to join the wind bow shock, implying that the gas in front of the motion of LMC~X--1 is first being highly ionized, before being swept up into the wind bow shock. We expect the wind bow shock to consist of highly photo-ionized gas, and indeed a clear, arc-like nebula structure is evident in the \HEII /\HB~ratio image of Figure 1(b) in \cite{2008ApJ...687L..29C}. This strongly photo-ionized arc resides between the ionization cone and the outer nebula. Although the outer nebula was argued to be shock-ionized \citep[possibly by a jet;][]{2007ApJ...667L.163C} there is not conclusive proof that it is entirely jet or photo-ionized. While the  region indicated energetic ionization, jet activity is not currently on in the system. In this work we investigate the wind bow shock scenario for the structure seen in Figure \ref{fig:radio}. 

The measurement of $D$ from the inner dashed lines in Fig. \ref{fig:radio} is a good estimate of the distance across the parabola of the wind bow shock. The 2.1 GHz radio emission appears to coincide with the \Halpha~nebula and overlap with the \HEII /\HB~arc. It could be due to bremsstrahlung (free-free) radiation from the shock front of the collisionally-excited gas \citep[e.g.][]{2005Natur.436..819G}, synchrotron radiation from within the jet-inflated region \citep[e.g.][]{2011ApJ...742...25Y}, non-thermal radio emission from relativistic particles in the wind bow shock \citep[e.g.][]{2010A&A...517L..10B,2016PhDT........76B}, or a combination of these.

\begin{table}
%\begin{minipage}{126mm}
\centering
\begin{tabular}{| c | c | c | c | c | }
\hline
D (pc) &  $R_{\rm o}$ (pc) & Error in D (pc) & Measurement\\
\hline
3.31 &0.96 & 0.8 & Din1\\
5.92  &1.71 & 0.8  & Dout1\\
\hline
\end{tabular}
\caption{An image taken of LMC (55 kpc away) has a conversion factor of 1.33 pc in space corresponding to 5 arc-seconds on the sky, so converting the measurements of the image shown in Figure \ref{fig:radio} we can estimate the diameter D of the bow shock. The error in measuring D is estimated to be about 0.8 pc. See text for details.}
%\end{minipage}
  \label{t:Ro}
\end{table}

Assuming that there is indeed a wind bow shock component to the nebula, from $D$ we get the standoff distance, $R_{\rm o}$, where the ram pressure of the wind and the ambient medium balance, which is given by Equation \ref{eqn:Wilkin}, and originates from \cite{1996ApJ...459L..31W}: 
\begin{eqnarray}
 R_{\rm o}=\sqrt{\frac{\dot{M}_{\rm w}v_{\rm wind}}{4\pi\rho_{\rm a}v^{2}_{\rm star}}}
 \label{eqn:Ro}
\end{eqnarray}
where $v_{\rm star}$ is the space velocity of the star. In this equation $\rho_{\rm a}$ is the density of the ISM, $\dot{M}_{\rm w}$ is the mass loss rate from the O8 star and $v_{\rm wind}=v_{\infty}$. The value we are interested in is $v_{\rm star}$, and inserting the other values in this formula, we measure a value of 0.638 pc for $R_{\rm o}$. This implies $R_{\rm o} = 2.95\times 10^{18}$ cm. The terminal wind velocity $v_{\rm wind}$ is assumed to be that for a normal O8 III star \citep{1990ApJ...361..607P,2009ApJ...697..573O}. We show our estimates of the density $\rho_{\rm a}$ of the ISM around LMC~X--1 in Table \ref{t:Ne}.

Rearranging Equation \ref{eqn:Ro} we see that the velocity of the star is given by:
\begin{eqnarray}
v_{\rm star} = \sqrt{\frac{\dot{M}_{\rm w}v_{\rm wind}}{4\pi\rho_{\rm a}R_{\rm o}^{2}}},
\label{eqn:vstar}
\end{eqnarray}
where $R_{\rm o}$ is estimated as shown in Table \ref{t:Ro}. Likewise if we know the system velocity with respect to the ambient medium we can provide a check on these calculations. Unfortunately, the LMC is too distant to obtain the proper motion of LMC~X--1. Its radial velocity (300 km s$^{-1}$) however, was obtained from the VLT/UVES spectrum as shown in Table \ref{t:measvel} and calculated in Section 3.

Approaching from the side of the measured densities of the ISM in this region, the average density of the ISM is about $\rho_{\rm a}=1.67\times 10^{-22}$ ${\rm g~cm}^{-3}$ assuming that the region is fully ionized and using a number density of $N_{\rm e}=40 {\rm cm}^{-3}$ \citep{pakullone}. With this density, a wind velocity of $v_{\infty}=2.6\times 10^{6}$ m s$^{-1}$ and $\dot{M}$ from \cite{2009ApJ...697..573O} we obtain the velocity of the star in its medium. The parameters that we are interested in are shown in Table \ref{t:Ne}.

\begin{table*}
\centering
%\begin{minipage}{126mm}
\begin{tabular}{| c | c | c | c | c | c |}
\hline
$\rho_{\rm a}$(${\rm g~cm}^{-3}$) &$R_{\rm o}$ (cm) & $v_{\rm wind}$ (km s$^{-1}$) at ISM & $\dot{M_{\rm o}} (M_{\odot}$/yr) \\
\hline

$1.67\times 10^{-22} $~\citep{1985AA...146..242B}		&$ 2.95\times 10^{18}$ &$2.6\times 10^{3} $~\citep{1990ApJ...361..607P}& $5\times 10^{-6} $~\citep{2009ApJ...697..573O}\\
$2.79\times 10^{-22} $~\citep{1964MNRAS.128..327F}	&$ 5.27\times 10^{18}$ &$1.4\times 10^{3} $~\citep{2009ApJ...697..573O}& $7.9\times 10^{-7}$~\citep{1989ApJS...69..527H}\\
%$1.62\times 10^{-21} $~\citep{1994AJ....107.1338W}	& n.a. &		n.a.		& $3.4\times 10^{-7} $~\citep{2001AA...369..574V}\\
%$2.68\times 10^{-21} $~\citep{2007ApJ...667L.163C}	&			n.a.					&	n.a.			& n.a. \\
%$6.69\times 10^{-23} $~\citep{pakullone}		&		n.a.						&	n.a. & n.a.	\\

\hline
\end{tabular}

\caption{Shown above are estimates of the wind bow shock standoff distance and parameters of interest from the literature in addition to the measured $R_{\rm o}$ found from Fig. \ref{fig:radio}. The citations for each literature value are given, showing the range in our parameters of interest. This range corresponds to a range of possible velocities for the LMC X-1 system.  The differences in the density of the ISM will give a variation in the velocity from Equation \ref{eqn:vstar}. We note that more extreme values are possible from other sources but have included only those with relatively low errors.}  \label{t:Ne}
% \end{minipage}
\end{table*}

Using the combination of parameters in Table \ref{t:Ne} we can then use Equation \ref{eqn:vstar} to explore the parameter range of the system velocity. These measurements are further complicated when the intricate density structure is taken into account. From Equation \ref{eqn:vstar} we see that if the density is one order of magnitude lower, the velocity estimates increase by a factor of three. 

Although there is a range of values shown in Table \ref{t:Ne} we can make a more precise estimate using what we know about the values shown. The estimate of $R_{\rm o} = 0.955~{\rm pc}=2.95\times 10^{18}~{\rm cm}$ for the standoff distance uses the best fit that we were able to achieve graphically. \cite{2009ApJ...697..573O} estimated the system parameters including the wind velocity and mass loss rate in a self-consistent way. Furthermore, using the density from \cite{1985AA...146..242B} we find a spatial velocity of:
\begin{eqnarray}
 v_{\rm star} \simeq 21.0 \pm 4.8 {~\rm km ~s}^{-1} 
\end{eqnarray}

To estimate the errors involved in this measurement we use the range of density, wind velocity, mass accretion and standoff radius shown in Table \ref{t:Ne} to be representative of the errors. We constrain ourselves to estimates of the density near our adopted value of $1.67\times 10^{-22}$ ${\rm g~cm}^{-3}$. Focusing on this scenario, the range of density from $1.67\times 10^{-22}$ ${\rm g~cm}^{-3}$ to $2.79\times 10^{-22}$ ${\rm g~cm}^{-3}$ we find a range of velocities from 21.0 km s$^{-1}$ to 16.4 km s$^{-1}$. For this scenario, this range gives us a rough estimate of the errors. We treat the error in our measurement of the velocity as $\delta v_{\rm star} = 4.8$ km s$^{-1}$. We note that it is possible to obtain more extreme values if larger ranges of density are considered, but in this case we restrict ourselves to the one scenario for simplicity. 

This is consistent with what we find for a range of densities. Using different values of the wind speed and $R_{\rm o}$ we find a most likely value around $v_{\rm star} \approx 21.0 \pm 4.8$ km s$^{-1}$. The measured radial velocities show a good agreement between the velocity of the emitting region and that of LMC~X--1, which strongly suggests these two are moving together, as we expect in the case of a bow shock.\footnote{We note that if we fit the wind bow shock to the outer diameter instead of the inner, we achieve a lower ($\simeq$7.63 km s$^{-1}$) velocity for the system.}

For the star to be moving supersonically, the spatial velocity must be greater than the sound speed, which was estimated at about 10 km s$^{-1}$ for the \HII~region. Notably as we go farther away from the bright O star the temperature and sound speed should decrease, but as noted by \cite{2005AJ....129..776N} there is the additional possibility of turbulence in the \HII~regions, which makes it hard to estimate what the sound speed might be farther out. It is possible that the sound speed is too high to have a wind bow shock at the low velocity predicted; however, making the assumption that the interstellar medium is neutral hydrogen, our $21.0 \pm 4.8$ km s$^{-1}$ would be supersonic. We conclude that our velocity range is consistent with supersonic motion and we adopt a value of $21.0 \pm 4.8$ km s$^{-1}$ as the spatial velocity of the star in the case of a wind bow shock inside the outer nebula structure. 

For a low space velocity we could produce a wind bow shock using the wind pressure of the star, but the velocity needed for a particular standoff distance changes with mass loss rate, wind rate and density. 

\section{The Origin of LMC~X--1}
\label{sec:origins}

We know that there are two types of runaway stars, those thrown out by dynamical ejection and those systems that received a kick velocity during the supernova explosion of the companion star in a binary \citep{1961BAN....15..265B}. The black hole companion of LMC~X--1 tells us that there was a supernova in the past that could have led to the latter scenario. Black holes can form by direct collapse and do not necessarily need a natal supernova with its associated recoil kick~\citep{2001ApJ...554..548F,2003Sci...300.1119M}.  If LMC~X--1 did get some kick velocity it could be quite low, as the formation of the black hole can potentially use up large amounts of material that would otherwise be able to be ejected during a supernova to give a kick velocity to the system. In this case, more mass would have had to be ejected (perhaps in winds) prior to supernova formation than was ejected in the supernova or used in the formation of the black hole.

Additionally, systems with a black hole can have high spatial velocities~\citep{2001Natur.413..139M,2002A&A...395..595M}, so a range of possibilities are present.\footnote{Recently~\cite{2017MNRAS.tmp...39R} found that a population model in which at least some black holes receive a (relatively) high natal kick best fits the observed distribution of black hole X-ray binaries with respect to the Galactic Plane.} Even in the case of a low velocity we will be able to investigate the evolutionary history of the system. We note that this type of kick should be limited to the plane of the binary orbit and would not be valid in the case of a pure jet powered nebula (after all, in a pure jet powered nebula, there would be no need to invoke a velocity to the system to explain the nebula structure). As we investigate the wind bow-shock scenario for the nebula it is possible that a small natal kick may have been given. As the system velocity is not very high, it is not a necessary condition that the velocity was created with a natal kick. It is equally possible that the velocity we see is the natural system velocity.

A runaway system can be traced back to its parent OB association if the spatial velocity and directionality can be determined \citep{2004RMxAC..21..128K}. If a parent OB association is determined then we can derive an age for the system (marking the time of the creation of the compact companion) by using the distance traveled and the velocity. The spatial velocity itself relates to the amount of mass lost during the supernova \citep{1999AA...352L..87N}. The age of the parent OB association should be roughly equal to the age of the binary system, so the turn-off mass at the time of the supernova should be the same as the initial mass of the primary. Combining this information will allow us to put constraints on the initial parameters of the progenitor of the HMXB as well as the evolutionary history of the system.

\begin{figure*}
\centering     
  \includegraphics[width=0.8\textwidth]{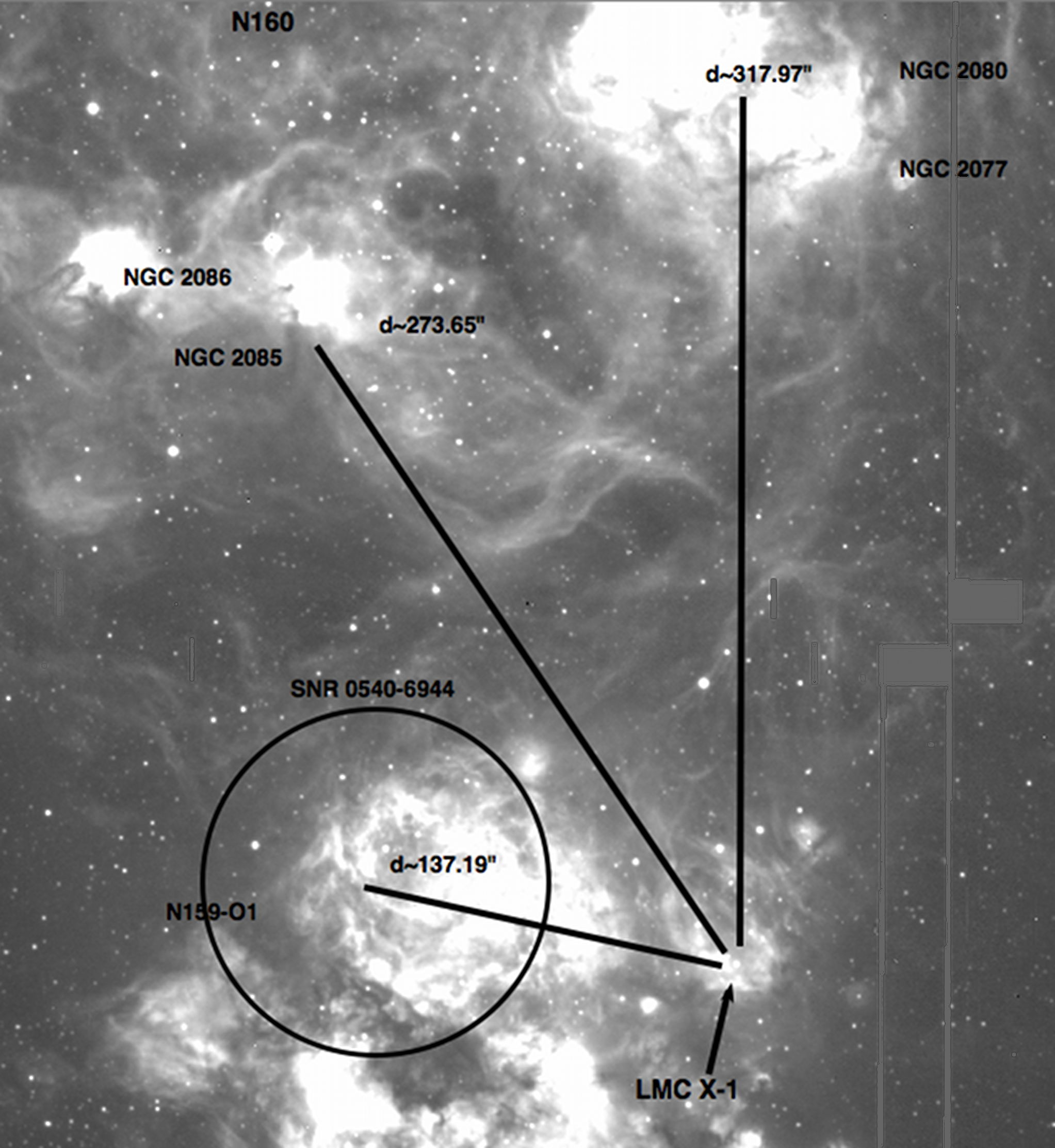} 
\caption{Wide-field WFI H$\alpha$ image of the nebulae and structure surrounding LMC~X--1 ($2 \times 500$ sec exposure). Several different clusters are labelled in this image as well as the supernova remnant SNR 0540-6944. The lines shown measure the distance $d$ and as a graphically measured distance it is therefore a minimum (using the 0.238"/pixel scale of the WFI). }  \label{fig:nebulaha}
\end{figure*}

If we adopt the hypothesis that LMC~X--1 is moving supersonically through the ISM we can use the bow shock shape to try to estimate the direction of its origin. We know from binary evolution theory that there are a few different scenarios that might have led to the system that we see today. Since we see a black hole that means that the originally most massive star has already had time to evolve through its lifetime and collapse. Before the collapse the originally most massive star could have transferred mass to its companion, and that companion would have then increased in mass to become the O star that we see today as Star 32. As the originally most massive star evolved, the core left over after mass transfer would collapse and form the black hole that we see today. If material was ejected a kick velocity provided to the system could lead to a high space velocity as we expect for a runaway system. We investigate this scenario and investigate one (of several) possible progenitor OB cluster for LMC~X--1.

LMC~X--1 is a wind accreting system \citep[e.g.][]{1993SSRv...66..309V}, i.e. the O star in the system is not currently experiencing Roche-Lobe Overflow (RLO). Furthermore the age of the O star is estimated at about 4 Myr by~\cite{2009ApJ...697..573O}. 

 We adopt a value of only $21.0 \pm 4.8$ km s$^{-1}$ for the spatial velocity of LMC~X--1 as discussed previously. If this velocity is creating a bow shock due to supersonic motion through the ISM that means that the parabolic shape of N\,159F is directional so we assume that the system had to come from one of the OB clusters shown in Figure \ref{fig:nebulaha}.

For a velocity of $21.0 \pm 4.8$ km s$^{-1}$ the travel time to several nearby clusters (indicated in Figure \ref{fig:nebulaha}) can be calculated. For a lifetime estimate of about 4 Myr the N\,159-O1 cluster seems to be a likely candidate, with a travel time of only $t\simeq1.2$ Myr, and placed near SNR 0540-6944 (coincidentally, as discussed in Section 5.3.1). From Figure \ref{fig:nebulaha}, the NGC 2086 and NGC 2085 groups would have travel times on the order of $t\simeq 2.4 Myr$, while NGC 2080 and NGC 2077 would have a travel time of approximately 2.8 Myrs. Although they all fit within a 4 Myr lifetime, we investigate the N\,159-O1 origin scenario.

This cluster is also valid for lower estimates of the velocity since it is quite nearby. Using the turn off mass of the N\,159-O1 cluster as approximately equal to the progenitor mass we will first investigate the evolution of the massive progenitor of the black hole and then we will use the system orbital parameters to see how much mass loss would have needed to occur if the velocity of $21.0\pm 4.8$ km s$^{-1}$ was due to a kick from the supernova that created the black hole in the system.

\subsection{Evolution of Star 32 and LMC~X--1}

We can place some restrictions on the progenitor of the black hole based on the evolutionary models of~\cite{2002RvMP...74.1015W} and \cite{2003ApJ...591..288H}. We note that according to those models, for a single massive star to be able to produce a core mass of around 10 $M_{\odot}$ the progenitor mass is roughly between 30 and 100 $M_{\odot}$ in the case of a solar metallicity star and zero metallicity star respectively (the zero metallicity case is purely theoretical so we assume that for our case of a low metallicity star in the LMC this general range should hold). Since we are looking in the LMC the star would be an intermediate case, but from the diagrams of single star evolution we see that in general the progenitor mass range above gives us a 10 $M_{\odot}$ black hole. In fact Woosley and Heger predict that even in a binary system with mass transfer for a progenitor mass larger than about 35 $M_{\odot}$ there is always going to be a neutron star or black hole produced in a supernova explosion.

This estimate corresponds with the model shown in~\cite{2002RvMP...74.1015W}. This model shows that it is possible to form a black hole either by fallback with a weak supernova or by direct collapse for a range of metallicities going from metal free all the way to solar for a progenitor mass above 25 $M_{\odot}$. There is some possibility that we can end up with a neutron star for certain equations of state but only near the higher metallicities, and since we are considering the LMC that particular possibility will not be considered.

Since the mass of the O star is currently thought to be $M_{\rm o} = 31.79$ $M_{\odot}$ \citep{2009ApJ...697..573O} that means that it has an expected lifetime of:
\begin{eqnarray}
\tau \approx \frac{1}{M^{p-1}}\times 10^{10}{\rm yr}
\label{eqn:tau}
\end{eqnarray}
where, for masses $M > 30$ $M_{\odot}$ we have $p = 3$ (if $M < 10$ $M_{\odot}$ then $p=4$), which gives 
\begin{eqnarray}
\tau = \frac{1}{M^{2}}\times 10^{10}{\rm yr} 
\label{eqn:lifetime}
\end{eqnarray}
for the more massive stars we are interested in. So in this case the O star in LMC~X--1 has a nuclear timescale of $\tau \approx 9.89$ Myr for evolutionary tracks between $25< M <40$ $M_{\odot}$. This is strong evidence that LMC~X--1 is still on the main sequence and therefore must be younger than the predicted $\tau \approx 9.89$ Myr lifetime. 

If the O star did accept mass from the previously most massive star in the system it may have experienced an effect known as rejuvenation due to the increase in mass (and fresh hydrogen). Basically this means that its lifetime estimate is starting from the point where it received the extra mass and not from its actual formation out of a molecular cloud. In this sense the lifetime estimate that we get for the O star is in fact the maximum time that has passed since rejuvenation. This time multiplied by the spatial velocity of the system would give an upper limit of how far the star could have traveled, i.e. $\tau \approx 9.89$ Myr would correspond to a distance of about 212 pc.

We can also place a lower limit on the amount of travel time the system has had by using the age estimate of~\cite{2009ApJ...697..573O}. The O star is currently near the 5 Myr isochrone for a 30 solar mass star. This indicates a current age of about 4.05 $-$ 4.22 Myr for the O star according to Orosz et al. (2009). This estimate is roughly consistent with LMC~X--1 coming from the region of N\,159-O1.

\subsection{Supernova in a Binary}
Following the analysis of \cite{1999AA...352L..87N} if we assume that the supernova took place in an initially circular orbit and that it was an instantaneous explosion, we can say that the position and velocity before and after the supernova are unchanged. Using conservation of mass we set the periastron velocity of the new orbit equal to the orbital velocity of the pre-supernova system and find:

\begin{eqnarray}
\frac{G(M_{\rm x}' + M_{\rm o})}{a_{\rm i}} = \frac{G(M_{\rm x}' + M_{\rm o} - \Delta M)}{a_{\rm f}}\times\frac{1+e}{1-e}
\end{eqnarray}
where $M_{\rm x}'$ is the mass of the pre-supernova star, $M_{\rm o}$ is the mass of the O star pre-supernova, which, in an instantaneous explosion, is the same as the mass of the O star afterwards. The mass of the core left over after the supernova will be taken as equal to $M_{\rm x}$. The radius for a circular orbit before is given by $a_{\rm i}$ and in this case $a_{\rm i}=(1-e)a_{\rm f}$ where $e$ is the eccentricity of the system after the explosion and $\Delta M$ is the mass lost from the system. Solving the above equation for the eccentricity we see that
\begin{eqnarray}
e = \frac{\Delta M}{M_{\rm x}' + M_{\rm o} - \Delta M}
\end{eqnarray}
after the supernova. The binary would be disrupted if $e >1$ or $\Delta M > (M_{\rm x}'+M_{\rm o})/2$. Since the mass transferred to the current O star obviously happened before the supernova explosion we assume that the O star mass before and after is roughly the same. We take values for the mass of the O star and the current mass of the black hole from \cite{2009ApJ...697..573O}  ($M_{\rm o} = 31.79$ $M_{\odot}$ and $M_{\rm x} = 10.91$ $M_{\odot}$ respectively). We can write the current mass of the system now as: 

\begin{eqnarray}
M_{\rm tot} = 10.91 + 31.79 M_{\odot} = 42.7 M_{\odot} = M_{\rm x}' + M_{\rm o} - \Delta M
\end{eqnarray} 
and since the black hole is the core left over after the massive star collapse this means that $M_{\rm x}' - \Delta M = 10.91$ $M_{\odot}$. Using the condition that the binary stays bound, $\Delta M < (M_{\rm x}'+M_{\rm o})/2$. Now we can solve for $\Delta M$. Substituting from above we see that $(2\times \Delta M - 31.79$ $M_{\odot}) < ((M_{\rm x}' = M_{\rm x}) + \Delta M)$ or $\Delta M < 42.7$ $M_{\odot}$. This means that the maximum mass that the system could have lost and still remain bound is $42.7$ $M_{\odot}$.

Losing mass from the binary during a supernova can impart velocity to the centre of mass of the system. We adopt a system velocity of $v_{\rm star} = v_{\rm s} = 21.0\pm4.8$ km s$^{-1}$. That being the case we can see how much mass would need to be lost from the system to create an intermediate spatial velocity of 21 km s$^{-1}$.

 Again following~\cite{1999AA...352L..87N} after a supernova explosion the velocity of the centre of mass of the binary will change by a value 

\begin{eqnarray}
v_{\rm s} = \frac{M_{\rm o}\times v_{\rm o}' - M_{\rm x}'\times v_{\rm x}' + \Delta M\times v_{\rm x}'}{M_{\rm x}' + M_{\rm o} - \Delta M}
\label{eqn:vel1}
\end{eqnarray}
where $v_{\rm o}'$ is the initial velocity of the O star in its orbit and $v_{\rm x}'$ is the initial velocity of the pre-supernova star. Since we have circular orbits beforehand we can simplify this equation even further by realizing that $v_{\rm o}' = \frac{2\pi a_{\rm o}'}{P'} = \frac{2\pi a_{\rm x}' M_{\rm x}'}{M_{\rm o}P'} = v_{\rm x}'(M_{\rm x}'/M_{\rm o})$ so the first two terms in Equation \ref{eqn:vel1} cancel and we are left with
\begin{eqnarray}
v_{\rm s} = \frac{\Delta M\times v_{\rm x}'}{M_{\rm x}' + M_{\rm o} - \Delta M} = e\times v_{\rm x}'
\end{eqnarray}

In this case $v_{\rm x}'$ is the velocity of the originally most massive star before the supernova. From Kepler's third law we know the period and semi-major axis relation and using physics for a binary system
\begin{eqnarray}
 v_{\rm x}'^{2} = \frac{4\pi^{2}}{P'^{2}}\frac{a'^{2}}{(1+(M_{\rm x}'/M_{\rm o}'))^{2}}
\end{eqnarray}
and, again using Kepler's law is re-written as
\begin{eqnarray}
 v_{\rm x}'^{2} = \frac{\pi}{a'}\frac{G\times M_{\rm o}^{2}}{M_{\rm o} + M_{\rm x}'}
\end{eqnarray}
since $M_{\rm o} = M_{\rm o}'$. Now, since the semi-major axis $a'$ is the initial value, $a'=a_{\rm i}=a/(1-e)$, and we can write $v_{\rm s}$ in terms of $e$, $M_{\rm o}$ and $M_{\rm x}'$,
\begin{eqnarray}
 v_{\rm s} = e\times \sqrt{\frac{\pi (1-e)}{a}\frac{G\times M_{\rm o}^{2}}{M_{\rm o} + M_{\rm x}'}}
\end{eqnarray}
therefore, if we know $\Delta M$ we can solve for the kick velocity $v_{\rm s}$. However, since we want to solve for $\Delta M$ we need a simpler formula. Following the work of~\cite{1999AA...352L..87N} this formula can be simplified to:

\begin{eqnarray}
 v_{\rm s} = 213\left[\frac{\Delta M}{M_{\odot}}\right]\left[\frac{M_{\rm o}}{M_{\odot}}\right]\left[\frac{P_{{\rm circ}}}{{\rm day}}\right]^{-1/3}\left[\frac{M_{\rm x}+M_{\rm o}}{M_{\odot}}\right]^{-5/3}\nonumber 
 \end{eqnarray}
(where units are in kms$^{-1}$) for a completely symmetric supernova explosion with a circular orbit of the pre-supernova system. In that case $P_{{\rm circ}}$ is the orbital period after re-circularization of the orbit, which is the period that we measure from the system today, i.e $P_{{\rm circ}} = P = 3.909$ days from~\cite{2009ApJ...697..573O}. Since we have an estimate for the velocity we rewrite this equation to solve for the mass lost during the supernova explosion and find

 \begin{eqnarray}
 \left[\frac{\Delta M}{M_{\odot}}\right] = 
 %\qquad\qquad\qquad\quad\qquad\qquad\qquad\qquad\qquad\nonumber 
 \left[\frac{v_{\rm s}}{213 {\rm kms}^{-1}}\right]\left[\frac{M_{\rm o}}{M_{\odot}}\right]^{-1}\left[\frac{P_{{\rm circ}}}{{\rm day}}\right]^{1/3}\left[\frac{M_{\rm x}+M_{\rm o}}{M_{\odot}}\right]^{5/3}
 \end{eqnarray}
 using $v_{\rm s} \approx 20$ km s$^{-1}$ and $\Delta M \approx 2.42$ $M_{\odot}$, and in the case where we consider the lower velocity (found by fitting the outer nebula structure) we find that even less mass needs to be lost from the system and $\Delta M \approx 0.85$ $M_{\odot}$, using $M_{\rm x}$ and $M_{\rm o}$ from \cite{2009ApJ...697..573O}.
 
This means that we do not have to eject very much mass from the system to create the range of kick velocities discussed in Section \ref{sec:windbowshock}. We also know that the progenitor mass $M_{\rm x}'$ should be approximately the turn off mass for the cluster where it originated from.

\subsection{Potential Origins}

Given the directionality of the system indicated through its bow shock structure and the velocity estimates of the system we estimate that, given the restrictions of our scenario, this system could have originated somewhere near the O star cluster N\,159-O1, which has an age of 10-30 Myr. Several other candidate origins for the LMC~X--1 system are shown in Figure \ref{fig:nebulaha} but we will take the nearby N\, 159-O1 nebula for this estimate. In particular, the combination of proximity and a local group of stars in the N\,159-O1 cluster that are the correct age for LMC~X--1 make this our current destination of choice (although, as mentioned in the beginning of this section, other origins are just as possible if you vary the density and system parameters sufficiently).  Since we have an age estimate from \cite{2009ApJ...697..573O} for LMC~X--1 of about 4.05 $-$ 4.22 Myr for the O star that puts a restriction on how far the system could have traveled (assuming rejuvenation of the current O star) since the supernova that created the black hole we see today. The region is complicated by many overlapping structures and there are younger clusters on top of the older population \citep{2005AJ....129..776N}. However, the region of N\,159-O1 has been surveyed by \cite{1998AAS..130..527T} and their work shows that the area of LH 101 (containing N\,159-O1) consists of two populations. The younger group is $\leq 2$ Myr old and the older group is between $3 - 6$ Myr in age.

We know that for nearby clusters, the turnoff age for the most massive stars will be the approximate age of the cluster. If this star did come from N\,159-O1 it probably would have had a mass somewhere around 60 $M_{\odot}$ if it was part of the older population. Using Equation \ref{eqn:tau} we see that the lifetime of a 60 $M_{\odot}$ star is $\tau \approx 3$ Myr. Additionally if the star has a velocity of about $21.0\pm4.8$ km s$^{-1}$ the travel time to the centre of N\,159-O1 is at least 1-2 Myr.

Conversely, if LMC~X--1 came from the direction of N160 (shown top left in Figure \ref{fig:nebulaha}), it could have potentially come from several different areas. We know that the current mass of the O star is $M_{\rm o} = 31.79$ $M_{\odot}$ so its lifetime is $\tau \approx 9.89$ Myr so this provides an upper limit on the distance the star could have traveled as shown in Figure \ref{fig:nebulaha}, since this is the lifetime for the star since rejuvenation. However, given the age estimate of LMC~X--1 from  \cite{2009ApJ...697..573O} we will rule out travel times greater than 6 Myr. Given the uncertainties involving the star velocity, we adopt N\,159-O1 as the most likely parent association for LMC~X--1.

\begin{table*}
\begin{center}
\begin{tabular}{ | c |}
\hline
LMC~X--1 System Parameters \& References \\ 
\hline
\end{tabular}
\begin{tabular}{| c | c | c | c |}
\hline
Quantity &  Value & Source \\
\hline
LMC~X--1 RA (J2000):	&05 39 38.7&\\
LMC~X--1 DEC (J2000):	&-69 44 36.0&\\
$L_{\rm o}$ & $3.17 \times 10^{5}$ L$_{\odot}$  & \cite{2009ApJ...697..573O}\\
$L_{\rm x}$ & $5 \times 10^{4}$ L$_{\odot}$ & \cite{1986Natur.322..511P}\\
Accretion Radius of Black hole &  $4.16\times 10^{6}$km & this work \\
Mass Accreted onto Black hole &  $1.30\times 10^{-8}M_{\odot}$/yr &  \cite{2009ApJ...697..573O}\\
$T_{\rm eff}$ & $33,200\pm 500$ K & \cite{2009ApJ...697..573O}\\
V magnitude	& 14.5	&  \cite{2009ApJ...697..573O}\\
%a  &   0.169 AU  &  this work\\
$M_{\rm o}$ &$ 31.79 \pm 3.48 M_{\odot}$ & \cite{2009ApJ...697..573O}\\
$M_{\rm x}$ & $10.91\pm 1.41 M_{\odot}$ & \cite{2009ApJ...697..573O}\\
$d_{\rm LMC}$ & 55 kpc & \cite{2008MNRAS.390.1762D}\\
$v_{\infty}$ & $2.6 \times 10^{3}$ km/s & \cite{1989ApJS...69..527H}\\
$v_{\rm wind}$ & $1.17\times 10^{6}$ m/s	&  \cite{2009ApJ...697..573O}\\
$K_{\rm opt}$ & $69.79 \pm 0.65$ km/s &  \cite{2009ApJ...697..573O}\\
$R_{\rm Star32}$ & $17.0 \pm 0.8 R_{\odot}$ &  \cite{2009ApJ...697..573O}\\
$P_{\rm orb}$ & $3.90917\pm 0.00005$ days & \cite{2009ApJ...697..573O}\\
$T_{\rm o}(\rm HJD~2,453,300+)$ & $91.3436\pm 0.0080$ days & \cite{2009ApJ...697..573O} \\
$\dot{M}$   &       $3.4 \pm 1.2 \times10^{-7}M_{\odot}$/yr &  \cite{2001AA...369..574V}\\
Velocity of LMC~X--1	& 21.0 $\pm 4.8$ km/s & this work\\
%$R_{\rm o}$ & ...1.94 pc & this work \\
$R_{\rm o}$(inner)	&0.638 pc	& this work\\
$R_{\rm o}$(outer)	&1.14 pc 	& this work\\
$[$OIII$]$5006.84$/$H$\alpha$	&0.8009	& this work	\\
\HEII~4685.71$/$H$\beta$		& 0.0542		& this work\\
$[$OIII$]$5007$/[$OI$]$6300	& 51.807	& this work	\\
H$\alpha/$H$\beta$			& 4.678	& this work\\
\hline
\end{tabular}
\end{center}
\caption{This table of system properties has been obtained primarily from the literature. $L_{\rm o}$ and $L_{\rm x}$ are the luminosities of the O star and its X-ray companion, respectively. The orbital period is $P_{\rm orb}$ and $T_{\rm o}$ is the time of inferior conjunction of the secondary. $T_{\rm eff}$ and the spectral type both refer to the massive O star and $M_{\rm o}$ and $M_{\rm x}$ are the respective masses of the O star and its black-hole companion. The distance to the system is approximately the distance to the Large Magellanic Cloud (LMC), $d_{\rm LMC}$.}
  \label{t:lmc}
\end{table*}

The mass of the progenitor of LMC~X--1 can be assumed to be approximately equal to the turn off mass of the older population of N\,159-O1, i.e.~$M \simeq 60$ $M_{\odot}$. Taking as an example the case where the star did indeed come from the N\,159-O1 region and had a progenitor mass of 60 $M_{\odot}$; this is roughly consistent with what we expect. For a single star we know from \cite{2002RvMP...74.1015W} that it may be possible for a star of this mass to produce a 10 solar mass black hole. We also know by following the analysis of \cite{1999AA...352L..87N} that in order to have a spatial velocity relative to the ISM of around 7 to 20 km s$^{-1}$ we would need to lose between 0.8 and 2 solar masses from the system. This means that a 60 solar mass star that leaves a 10.3 solar mass core and ejects 2 solar masses out of the system would need to transfer 48 solar masses to its companion (however, there was most probably a large loss of mass due to stellar wind as well). Since the current mass of the O star is thought to be about $31.79$ $M_{\odot}$ there would have had to be additional mass loss from the system. From \cite{2002RvMP...74.1015W} massive stars can lose a lot of mass due to their strong winds. There is also the possibility that the system did not have a conservative mass transfer but rather went through a phase of non-conservative mass transfer (this can result in a spiraling-in of the system and additional mass loss), and so a 60 solar mass star would be a reasonable progenitor for the LMC X--1 system. Common envelope evolution can occur in binaries which end up forming black holes and support the larger mass loss case. We note that this is subject to modeling but we will consider that LMC~X--1 could have had a 60 $M_{\odot}$ companion as its originally most massive star in our scenario.

\subsubsection{SNR 0540-6944}

Since we have a black hole binary system the originally most massive star could have ended its life with a supernova explosion. We can see by the directionality indicated by the bow shock and the current velocity estimates of the system that LMC~X--1 may have originated in the direction of the cluster N\,159-O1 or the more distant NGC 2077, and NGC 2085. The bow shock gives an acceptable indicator of the direction this system is traveling and, following the line of sight back, an interesting supernova remnant (SNR), SNR 0540-6944 (at an RA of 05:40:05 and DEC -69:44:07), was discovered quite close by. This discovery was made by \cite{1997PASP..109..554C} and the remnant is at least  36 pc away from the LMC~X--1 system. This is the minimum distance we can have since it corresponds to the line drawn from one to another on the sky (i.e. since LMC~X--1 is at 05:39:38.7 -69:44:36 they appear to have a separation of 98 arc-seconds and assuming a distance of 55 kpc to the LMC this means a minimum separation of 36 pc). 

However, if we compare the velocity of the SNR with LMC~X--1 we find a discrepancy. The SNR has a fast expanding shell with velocity offsets of $-180$ to $+130$ km s$^{-1}$ as compared to the background \HII~region \citep{1997PASP..109..554C} while LMC~X--1 has a velocity between 20 and 60 km s$^{-1}$ depending on density estimates. The high velocity of this shell coupled with the fact that LMC~X--1 is currently located outside the radius of the SNR as shown in Figure \ref{fig:nebulaha} also excludes a causal relation between the two objects.

This is further confirmed by the travel times involved, since 1 pc $\approx 3.086 \times 10^{16}$ m that means that if the star is moving about $21.0 \pm 4.8$ km s$^{-1}$ it will take $1-2$ Myr for the star to have traveled 36 pc to the centre of the SNR.

For us it allows a correspondence between the two objects to be ruled out. The velocity of $21.0\pm 4.8$ km s$^{-1}$ as measured from Table \ref{t:measvel} combined with:
\begin{eqnarray}
t_{\rm snr}=\frac{d_{\rm snr}}{v_{\rm star}}
\end{eqnarray}
where $d_{\rm snr}$ is the estimated distance and $v_{\rm star}$ in this case is the measured velocity of the LMC~X--1 system relative to the ISM that surrounds it. This gives a travel time of $t_{\rm snr}=1-2$ Myr. This is much greater than the maximum age of the SNR, which is $2\times 10^{4}$ years \citep{2000ApJ...536L..27W}.

The LMC~X--1 environment shown in H$\alpha$ in Figure \ref{fig:nebulaha} includes the supernova remnant SNR 0540-6944 whose radius is outlined nearby. This is an extremely active region and the direction of the bow shock indicates that LMC~X--1 is traveling away from the region of its SNR neighbor while the velocity range of LMC~X--1 excludes a causal relation between the two. As discussed above, this is because the upper limit on SNR age is 0.02 Myr and in that time LMC~X--1 could not have traveled the distance needed.

\section{Summary and Conclusions}
\label{sec:concl}

Our new observations have led to a new summary of parameters in this paper, as shown in Table \ref{t:lmc}. We have presented new optical spectra with UVES on the VLT and narrowband images with the WFC on the 2.2 m ESO/MPI telescope of LMC~X--1 and its nebula in proximity to the X-ray binary. For the optical spectrum we use a new spectrum extraction technique to obtain both radial velocities and fluxes \citep{2014PASP..126..170R}. We confirm the spectral type of the O-star and show some evidence that the exact spectral type is likely Of but we cannot rule out Of?p.

We present new ATCA 2.1 GHz radio continuum imaging showing the arc-like nebula around LMC~X--1 that is tracing the shocked gas. We investigate the wind bow shock scenario for the nebula structure found around LMC~X--1. We compare the radio emission to existing WFI imaging of the area in \Halpha and fit a parabolic shape to the inner arc to estimate the parameters of a wind bow shock structure. This wind bow shock, bright in He\,{\sc ii}, can be represented as a separate structure to the large-scale collisionally ionized nebula. The nebulosity around LMC~X--1, visible in numerous emission lines, can be explained by an ionization cone \citep{2008ApJ...687L..29C}, a wind bow shock formed by the O star wind and a larger, shock-ionized nebula powered kinetically, possibly by the jets of LMC~X--1. The radio emission could be non-thermal emission from the wind bow shock, or bremsstrahlung (or synchrotron) emission associated with the jet inflated nebula. For either wind or jet-powered origins, this would represent one of the first radio detections of such a structure.

From our VLT/UVES spectrum we detect nebular emission; these emission lines include the ionized region plus likely some of the bow shock. By fitting a wind bow shock structure to the inner arc nebula around LMC~X--1 we derive a space velocity of $v_{\rm star} \approx 21.0 \pm 4.8$ km s$^{-1}$, indicating possible supersonic motion through an ISM of neutral H. This velocity excludes the possibility that LMC~X--1 could be causally related to the nearby supernova remnant SNR 0540-6944. The velocity is consistent with the radial velocity measurements taken directly from the spectrum of LMC~X--1, which shows that the emission from the shocked region is moving with the star.

We constrain the positional origin of LMC~X--1 using its now refined space velocity, and assess the scenario of its birth within a known Supernova Remnant (SNR). We investigate N\,159-O1 as a possible origin for the LMC~X--1, but we do not rule out other nearby regions. In the scenario of an N\,159-O1 origin, we suggest a progenitor mass of $\sim 60$ M$_{\odot}$ for the black hole.

\section{Acknowledgements}
%\acknowledgements
We would like to acknowledge many valuable conversations with Ryan Cooke about his work on LMC X-1. We would also like to acknowledge the support of several colleagues, namely: Richard McDermid, Ray Norris, Warren Reid, David Frew, Stuart Ryder and Quentin Parker. DMR wishes to thank Rob Fender and Elena Gallo for useful discussions about nebulae powered by X-ray binary jets. We used the karma and miriad software package developed by the ATNF. The Australia Telescope Compact Array is part of the Australia Telescope, which is funded by the Commonwealth of Australia for operation as a National Facility managed by CSIRO. 

%%%%%%%%%%%%%%%%%%%%%%%%%%%%%%%%%%%%%%%%%%%%%%%%%%



\bibliographystyle{aasjournal}
\bibliography{mainlmc2}
%%%%%%%%%%%%%%%%%%%%%%%%%%%%%%%%%%%%%%%%%%%%%%%%%%

%%%%%%%%%%%%%%%%% APPENDICES %%%%%%%%%%%%%%%%%%%%%

%\appendix

%%%%%%%%%%%%%%%%%%%%%%%%%%%%%%%%%%%%%%%%%%%%%%%%%%
\allauthors

% Don't change these lines
%\bsp	% typesetting comment
%\label{lastpage}
\end{document}